\documentclass[preprintnumbers,showpacs,amsmath,amssymb,floatfix,9pt,prd,twocolumn,
superscriptaddress,nofootinbib]{revtex4}
\usepackage{graphicx}
\usepackage{latexsym}
\usepackage{epsfig}
\usepackage{color}
\usepackage{amssymb}
\usepackage{hyperref}
\hypersetup{colorlinks,citecolor=blue}

\begin{document}

\title{Decoupling gravitational sources by MGD approach in Rastall gravity}

\author{S. K. Maurya}
\email{sunil@unizwa.edu.om} 
\affiliation{Department of Mathematics and Physical Science, College of Arts and Science, University of Nizwa, Nizwa, Sultanate of Oman}

\author
{Francisco Tello-Ortiz}
\email{francisco.tello@ua.cl}
\affiliation{Departamento de F\'isica,
Facultad de ciencias b\'asicas, Universidad de Antofagasta, Casilla 170, Antofagasta, Chile}

\begin{abstract}
In the present work, we investigate the possibility of obtaining stellar interiors for static self-gravitating systems describing an anisotropic matter distribution in the framework of Rastall gravity through gravitational decoupling by means of minimal geometric deformation approach. Due to Rastall gravity breaks down the minimal coupling matter principle, we have provided an exhaustive explanation about how Israel-Darmois junction conditions work in this scenario. Furthermore, to obtain the deformed space-time, the mimic constraint procedure has been used. In order to check the viability of this proposal, we have applied it to the well known Tolman IV solution. A complete thermodynamic description of the effects introduced by the additional source is given. Additionally, the results have been compared with their similes in the picture of pure general relativity, pure Rastall gravity and within the framework of general relativity including gravitational decoupling. To perform the mathematical and graphical analysis we have taken the gravitational decoupling constant $\alpha$ and the Rastall's parameter $\lambda$ as free parameters and the compactness factor describing the general relativity sector to be $0.2$. Besides, to provide a more realistic picture we have bounded both parameters $\alpha$ and $\lambda$ by using real observational data to explore the limits of the theory under this particular model. Applications to study neutron or quark stars are suggested by using this methodology. 

\end{abstract}
\maketitle

\section{Introduction}
Put forward {by} P. Rastall in 1972 \cite{r1}, the so-called Rastall gravity theory can be seen as a modified gravity theory or as a generalization of Einstein gravity theory \i.e General Relativity (GR from now on). Rastall's proposal is based on the argument that the energy-momentum tensor, which fulfills the conservation law (null divergence) in a flat space-time (Minkowskian space-time), does not necessarily fulfill it in a curved background. The violation of Bianchi's identity occurs through the introduction of a covariant term in the Einstein field equations through a dimensionless coupling constant $\lambda$. Specifically, this term corresponds to the Ricci's scalar curvature R. Although the field equations given by Rastall do not have an associated Lagrangian density from which they can be obtained, these as a generalization of the GR equations respect the symmetries of the theory \i.e, the general coordinate transformation. 

Despite that the new term is introduced by hand, its prevalence modifies not only the field equations but also the way of coupling material fields to the gravitational interaction. Clearly, the principle of minimal coupling matter breaks down. However, this brings with it new and intriguing contributions which can be useful to understand certain commonly studied phenomena such as cosmological issues, collapsed structures (black holes), stellar structures, gravitational waves, etc. In this direction, Rastall gravity is as competitive as other modified gravity theories such as f(R) and f(R,$\mathcal{T}$) theories. As it is well known f(R) gravity theory was developed to address inflationary cosmological problems \cite{r2,r3}. Nowadays, f(R) gravity has been used in a wide context by addressing cosmological problems such as the existence of dark components (dark energy and dark matter), stellar structures, among others. An incomplete list of recent works concerning these issues can be found at the following references \cite{r4,r5,r6,r7,r8,r9,r10,r11,r12,r13,r14,r15,r16,r17,r18,r19,r20,r21,r22} (and references contained therein). Furthermore, to face related open questions in the cosmological scenario f(R,$\mathcal{T}$) gravity is also a promising approach \cite{r23}. In this respect f(R,$\mathcal{T}$) gravity theory can be seen as an extension of f(R) gravity. Where $\mathcal{T}$ stands for the trace of the energy-momentum tensor. The corrections coming from the trace of the energy-momentum tensor can be attributed to quantum effects \cite{r23}. A wide variety of works available in the literature devoted to tackle the accelerated expansion of the Universe, energy conditions, stellar interiors and so on, are found at the following references \cite{r24,r25,r26,r27,r28,r29,r30,r31,r32,r33,r34,r35,r36,r37,r38,r39,r40,r41,r42,r43,r44,m1,m2,m3,m4}. In comparing f(R,$\mathcal{T}$) gravity with Rastall theory, both theories break the minimal coupling matter scheme. However, the former smashes the minimal coupling matter principle by introducing matter and geometric terms (curvature invariants) while the second one by inserting only geometric objects, precisely the Ricci's scalar. The effects introduced by the additional term have been extensively studied on different fronts to test at least theoretically how close the results are in comparison with the broad support that GR has. In this respect, the well known Tolman solutions have been extended into the Rastall gravity arena \cite{r45} in order to contrast the behavior of the main salient features with the corresponding GR ones. Furthermore, to reinforce the study of stellar interiors in the Rastall scenario an anisotropic model in the background of Krori-Barua (KB) space-time was done in \cite{r46} (in \cite{r47} the same authors considered KB space-time supplemented by an equation of state, specifically a quintessence model) and an isotropic compact object using conformal Killing vector technique was reported in \cite{r48}. Regarding the strong gravitational regime, remarkable solutions of black holes were presented in \cite{r49,r50}. These works motivate the investigation of the most exciting features and properties of these fascinating objects. For example the thermodynamic was studied in \cite{r51,r52,r53} and rotating black holes were addressed at references \cite{r54,r55}. It is worth mentioning that in the context of black holes solutions GR and Rastall gravity share the same vacuum solution \cite{r1}. This is so because when the Rastall parameter goes to zero ($\lambda\rightarrow 0$), the Rastall contribution disappears. Moreover, in the limit $\lambda\rightarrow 0$ all GR results are recovered.  

Based on these good antecedents, in the present work, we want to investigate the possibility of obtaining compact structures which serve to describe stellar interiors in the light of Rastall gravity theory. To accomplish it we employ gravitational decoupling by means of minimal geometric deformation (MGD) approach \cite{r56,r57}. This methodology was developed to deform Schwarzschild space-time \cite{r58,r59} in the Randall-Sundrum framework \cite{r60,r61}. In few words, gravitational decoupling through MGD algorithm consists in to extent well behaved isotropic solutions (however is not necessary to consider the input solution to be isotropic, it can be anisotropic, charged, etc.) of Einstein gravity theory to anisotropic domains. To do so, one needs two main ingredients: i) add an extra source $\theta_{\mu\nu}$ to the energy-momentum tensor of the seed solution via a dimensionless coupling constant $\alpha$. The presence of this extra piece in conjunction with a spherically symmetric and static space-time (in Schwarzschild like-coordinates) leads to an intricate system of equations with seven unknown functions (if the seed solution is taking to be isotropic). To solve this complicated system and translate the isotropic fluid distribution to an anisotropic scenario ii) perform the MGD on one of the metric potentials (usually on the radial metric component $e^{\lambda}$). With this deformed potential in hand, the tangled system is separated into two simpler sets. The first one is the usual Einstein system and the second one contains the $\theta$-sector and the decoupler function $f(r)$ introduced in the MGD to split the system of equations. However, the latter one contains four unknown and only three equations. Then this system must be supplemented with extra information in order to close the problem. At this stage, a couple of comments are in order. First, after decoupling the systems, the resulting ones fulfill Bianchi's identities (the conservation law of {their} corresponding energy-momentum tensors), meaning that the original source and the extra one only interact gravitationally. Second, the additional term could represent a scalar, vector or tensor field. Moreover, in principle, this new sector could not necessarily be described by GR. For a more detailed discussion of how this machinery works see sections \ref{sec3} and \ref{sec4} and the following references 
\cite{r62,r63,r64,r65,r66,r67,r68}.

This last two years gravitational decoupling using MGD has attracted many adepts. In this respect some well known solutions (uncharged, charged) of the Einstein field equations have been extended by using MGD \cite{r69,r70,r71,r72,r73,r74,r75,r76,r77,r78,r79,r791,r792}. Also, Black holes in
$3+1$ (Schwarzschild outer space-time) \cite{r80} and $2+1$ (BTZ black hole) \cite{r81} dimensions were extended, the  (anti) de Sitter space-time was worked in \cite{r82} and also the inverse problem was addressed in $3+1$ dimensions \cite{r83} and $2+1$ dimensions including cosmological constant \cite{r84}. Regarding another branches the method has been employed in cloud of strings \cite{r85}, Klein-Gordon scalar fields as an extra matter content \cite{r86}, extended to isotropic coordinates \cite{r87} and ultra compact Schwarzschild stars, or gravastars \cite{r88}. Moreover, the method was widespread to include deformation on both metric potentials, it was called the extended-MGD \cite{r89,r90,Maurya19}. More recently gravitational decoupling was used to investigate higher dimensional compact structures \cite{r91} and spread out to the context of Lovelock \cite{r92} and f(R,$\mathcal{T}$) \cite{r93} gravity theories and in the cosmological scenario \cite{r94}.

So, as we pointed out above our main goal is to introduce gravitational decoupling by means of MGD in the Rastall gravity picture. This will bring new insights on how compact objects behave by the inclusion of local anisotropies in the light of Rastall theory. Moreover, the possibility of comparing with other modified gravity theories and GR results in order to check the viability from both theoretical and experimental point of view. To do so, we have followed the same procedure given in \cite{r57}. The approach proposed in \cite{r57} in order to solve the $\theta$-sector is to impose some suitable restrictions on the thermodynamic quantities that characterize the seed solution (the isotropic pressure $p$ and the energy-density $\rho$) and the components of the extra source $\theta_{\mu\nu}$. These restraints are referred as mimic constraints. These mimic constraints yield to an algebraic or differential equation that allows to obtain the deformation function $f(r)$. Each mimic constraint leads to a different anisotropic solution, for example, the two common mimic constraints worked in the literature are: i) $p=\theta^{r}_{r}$ and ii) $\rho=\theta^{t}_{t}$. This means that the radial component of the $\theta$-sector mimics the isotropic pressure the temporal one mimics the energy-density of the seed solution. Although there {is not} a physical foundation to support the aforementioned choices, until now they have not presented any physical or mathematical inconsistency or any behavior that is detrimental to what is reported within the framework of general relativity. Moreover, studies conducted in general relativity concerning to interior solutions with anisotropic component have been favored (or reinforced) in a certain way when the described mechanism has been considered. However, in favor of these considerations, it should be noted that the mimic constraints have been imposed at the level of the equations of motion, which ensures the closure of the system of equations to be solved and also a correct physical and mathematical behavior of the $\theta$-sector components. Therefore, {one} ends with a well-behaved solution. In addition, the virtues of each restriction are unique. For example, when the $r-r$ component of the $\theta$-sector mimics the seed pressure $p(r)$, the total mass of the structure does not change, it is only redistributed inside the object. On the other hand, the election $\rho=\theta^{t}_{t}$ changes the total mass of the compact object. 
Consequently, the first case does not distinguish between an isotropic and an anisotropic object of similar characteristics (mass and radius). This is so because the surface gravitational red--shift $z_{s}$ is the same in magnitude. Nevertheless, in the second case, one can differentiate between a distribution with isotropic material content from another with anisotropic content, since in the latter $z_{s}$ varies as expected with respect to its isotropic counterpart, changing its magnitude. It is worth mentioning that another way to face the problem is imposing and adequate form of the decoupler function $f(r)$ respecting all the physical and mathematical requirements as was done in \cite{r69,r70,r91}.

In the present work, we have considered both mimic constraints. Due to Rastall gravity departs from GR only by the presence of the Ricci scalar coupled to the theory through the so-called Rastall's parameter $\lambda$, then there are not higher-order derivative terms of the metric potentials (no more than two spatial derivatives) in the theory. This feature facilitates the gravitational decoupling and we have done it in such a way that the deformation function and the resulting components of $\theta$-sector contain the effects of Rastall gravity. This allows the new contributions to be compared exhaustively with respect to what is obtained in pure Rastall gravity and the results of RG and RG + MGD. In addition, to contrast our results we have taken the geometry of the inner space-time to be Tolman IV. This solution which describes a spherically symmetric and static object whose material content respects a perfect fluid distribution has already been extended to anisotropic domains by MGD \cite{r57} and has also been studied in the Rastall gravity picture \cite{r45}. One of the most notable features of Rastall's theory of gravity is that any perfect fluid solution of Einstein's equations is also a solution of it. Obviously, this is from the geometrical point of view because the material content is different due to Rastall contribution. Furthermore, to handle with the numerical part we have taking as free parameters the gravitational decoupling constant $\alpha$ and the Rastall's parameter $\lambda$ and considering a compactness factor within the allowed range for compact stars to be $u=m_{GR}(R)/R=0.2$. In this respect, it is worth mentioning that the coupling constant $\alpha$ plays an important role in the behavior on the main salient features and has a relevant incidence on the mass of the compact structure. Although we have only considered positive values for $\alpha$, in order to obtain a physically acceptable solution. Notwithstanding, $\alpha<0$ is not completely discarded. Particularly the choice $\rho=\theta^{t}_{t}$, for some solutions such as Heintzmann IIa \cite{r71} and Durgapal-Fuloria \cite{r73}, {requires} that $\alpha$ takes negative values to avoid non-physical {behaviors}. In our case,  $\alpha<0$ is forbidden in considering both mimic constraints. So, the final deformed Tolman IV space-time is in complete agreement with what is expected in this research field. On the other to check the feasibility and limits of the theory we have bounded both $\alpha$ and $\lambda$ by using real observational data. 
Finally, we want to mention that it is the first time that gravitational decoupling by MGD is used in the Rastall gravity scenario. 

So, the article is organized as follows: In Sec. \ref{sec2} we revisited in brief the main ingredients of Rastall gravity theory and {its} comparison with other theories. In Sec. \ref{sec3} the field equations for multiple sources are presented. Sec. \ref{sec4} discuss the gravitational decoupling via minimal geometric deformation scheme in the Rastall gravity framework. Next, in Sec. \ref{sec5} the matching conditions are analyzed and extensively discussed as well as the binding energy in present context. In Sec. \ref{sec6} Tolman IV space-time is projected into the anisotropic domain using the methodology previously discussed. Furthermore, the problem is faced by employing the mimic constraint approach, in order to determine the decoupler function $f(r)$ and the new sector, the $\theta_{\mu\nu}$ one. Sec. \ref{sec7} talks about the physical implications of the mimic constraints procedure on the principal macro physical observables of the model as well as the maximum and minimum order of magnitude of the free parameters $\alpha$ and $\lambda$ allowing to contrast with real values of compact objects. In the following section, Sec. \ref{sec8} the principal physical, observational and mathematical implications of the obtained model are highlighted. Finally, Sec. \ref{sec9} provides some remarks of the study reported in this article.   

\section{Revisiting in short: Rastall gravity theory}\label{sec2}

The main idea behind Rastall's proposal \cite{r1} is to abandon the free divergence of energy-momentum tensor in a curved space-time. Explicitly it reads
\begin{equation}\label{eq1}
\nabla_{\mu}T^{\mu}_{\nu}\neq 0. 
\end{equation}
So, this non-conserved stress-energy tensor introduces an unusual non-minimal coupling between matter and geometry. Specifically, this non-minimal coupling is carried out into the theory by the following assumption on the divergence of the energy-momentum tensor
\begin{equation}\label{eq2}
\nabla_{\mu}T^{\mu}_{\nu}=\lambda \nabla_{\nu}R.    
\end{equation}
In the above expression $R\equiv g^{\mu\nu}R_{\mu\nu}$ stands for the Ricci's scalar and $\lambda$ is the so-called Rastall's parameter which is used to depict the distraction from GR and measures
the affinity of the space-time geometry to couple with matter field in a non-minimal fashion. The assumption given by Eq. (\ref{eq2}) is completely consistent with the following field equations 
\begin{equation}\label{eq3}
R_{\mu\nu}-\frac{1}{2}Rg_{\mu\nu}=\kappa_{R}\left(T_{\mu\nu}-\lambda R g_{\mu\nu}\right),  
\end{equation}
where $\kappa_{R}$ is the Rastall gravitational constant. It is worth mentioning that in the limit $\lambda\rightarrow 0$ Einstein's field equations are recovered, then the energy-momentum tensor (\ref{eq2}) is conserved. However, the above field equations (\ref{eq3}) can be expressed in the following form
\begin{equation}\label{eq4}
G_{\mu\nu}=\kappa_{R}T^{(\text{eff})}_{\mu\nu},   
\end{equation}
and in some sense one {regains} the standard result $\nabla^{\mu}T^{(\text{eff})}_{\mu\nu}=0$.
Now, by taking the trace of Eqs. (\ref{eq3}) the Ricci scalar can be written  as 
\begin{equation}\label{eq5}
R=\frac{\kappa_{R}T}{4\lambda\kappa_{R}-1},   
\end{equation}
then the effective stress-energy tensor reads
\begin{equation}\label{eq6}
T^{(\text{eff})}_{\mu\nu}=T_{\mu\nu}-\frac{\gamma T}{4\gamma-1}g_{\mu\nu},
\end{equation}
where $\gamma=\kappa_{R}\lambda$. From now on we shall assume $\kappa_{R}=1$, then $\gamma=\lambda$. From Eq. (\ref{eq6}) one can infer several constraints. Again taking $\lambda=0$ the Rastall sector disappears and one recast GR. If a traceless energy-momentum tensor is considered, such as the electromagnetic one, Rastall contribution is totally vanished because $T=0$. Additionally, $\lambda=1/4$ represents non-physical situation. So, this value must be excluded in order to avoid inconsistencies. From now on, we will consider $T_{\mu\nu}$ to be a perfect fluid matter distribution, which is given by
\begin{equation}\label{eq7}
T_{\mu\nu}=\left(\rho+p\right)u_{\mu}u_{\nu}-pg_{\mu\nu}.
\end{equation}
We utilize a comoving fluid $4$-velocity $u^{\sigma}=e^{-\nu/2}\delta^{\sigma}_{t}$, and $\rho$ and $p$ are representing the energy-density and the isotropic pressure respectively. So, the components of $T^{(\text{eff})}_{\mu\nu}$ are given by
\begin{equation}\label{eq8}
{T^{t\ (\text{eff})}_{t}}=\rho^{(\text{eff})}= \frac{\left(3\lambda-1\right)\rho+3\lambda p}{4\lambda-1},
\end{equation}
\begin{eqnarray}
 \label{eq9}
{T^{r\ (\text{eff})}_{r}=T^{\varphi\ (\text{eff})}_{\varphi}=T^{\phi \ (\text{eff})}_{\phi}}=-p^{(\text{eff})}=- \frac{\left(\lambda-1\right)p+\lambda \rho}{4\lambda-1}. 
\end{eqnarray}

In obtaining the expressions (\ref{eq8}) and (\ref{eq9}) we have used $T=\rho-3p$. Moreover, as before $\lambda=0$ yields us to Eq. (\ref{eq7}). As we will see later the isotropic quantities $\rho$ and $p$ will be separated carrying out in their expressions the corresponding Rastall contributions. 

Despite its attributes, recently was claimed by Visser \cite{visser} that Rastall gravity theory is equivalent to Einstein's general relativity theory. The main concern of Visser was that the defined energy-momentum-tensor provided by Rastall was not
right and that Rastall's theory is just the rearrangement of the matter sector of GR. Of course as can be seen from Eq. (\ref{eq4}) the fields equation given in the original Ratsall's article \cite{r1} can be adjusted to recast the usual Einstein field equations. However, in distinction if the energy-momentum tensor is conserved or not, this \emph{rearrangement} can be performed in any modified gravity theory e.g. $f(R,\mathcal{T})$ gravity \cite{r23}, $f(R)$ theory \cite{felice} among others, where the terms non conforming Einstein tensor are grouped giving rise to an effective energy-momentum tensor. So, it does not imply that GR is equivalent
to these theories. In fact, such  equivalence exists only in particular cases such as: Putting $f(R)=R$ in $f(R)$ gravity theory, or dropping out the $\mathcal{T}$ term and setting $f(R)=R$ in $f(R,\mathcal{T})$ gravity \cite{epjc}. Therefore, one can conclude that Rastall's proposal is not equivalence to GR unless $\lambda\rightarrow 0$. What is more, recently, from the $f(R,\mathcal{T})$ Lagrangian formulation was obtained the corresponding Lagrangian functional associated with Rastall's theory \cite{moraes} and G\"{o}del-type solutions in the cosmological scenario were investigated \cite{moraes1}.
At this point it is relevant to compare Rastall's theory with other modified theories of gravity which violate the conservation of the energy-momentum tensor. As it is well known the original $f(R)$ gravity proposal respects Bianchi's identities. However, this theory was extended by  including extra terms which violate the minimal coupling matter principle. Then, the energy-momentum tensor associated to this formulation in not conserved \cite{bertolami}. Specifically, the modified Einstein-Hilbert action reads
\begin{equation}\label{new1}
S=\int\bigg\{\frac{1}{2}f_{1}(R)+\left[1+\beta f_{2}(R)\right]\mathcal{L}_{m}\bigg\}\sqrt{-g}d^{4}x,    
\end{equation}
where $f_{1}$ and $f_{2}$ are arbitrary {functions} of the Ricci's scalar and $\mathcal{L}_{m}$ is the Lagrangian matter characterized by the constant $\beta$. The field equations provided by the above action (\ref{new1}) are given by
\begin{equation}\label{new2}
\nabla^{\mu}T_{\mu\nu}=\frac{\beta F_{2}}{1+\beta f_{2}}\left[g_{\mu\nu}\mathcal{L}_{m}+T_{\mu\nu}\right]\nabla^{\mu}R.
\end{equation}
As it is observed, Eqs. (\ref{eq2}) and (\ref{new2}) are in some sense comparable, since both {shown} a non-trivial coupling between the gravitational and material sectors. In Eq. (\ref{new2}) $T_{\mu\nu}$ is representing the usual energy-momentum tensor describing isotropic, anisotropic matter distribution, etc. If the term $\beta f_{2}$ is taking to be constant, namely $\beta f_{2}=K$ then $F_{2}\equiv df_{2}/dR=0$, thus the usual conservation law is regained. Also the familiar conservation equation is recovered from (\ref{new2}) when $\beta\rightarrow 0$ as in the Rastall case when $\lambda\rightarrow 0$.

The theory described by (\ref{new1}) has been analyzed and contrasted with the well established GR results, in several situations such as, in the study of stellar interiors \cite{bertolami1}, Newtonian approximation \cite{bertolami2} and galaxy/cluster dynamic \cite{bertolami3}. Respect to stellar interiors one of the most important concern in this area is the equilibrium of the compact structures under different forces. This analysis is carried out by employing the modified Tolman--Oppenheimer--Volkoff (TOV) equation. The TOV equation in the GR context reflects the conservation of the energy-momentum tensor. Of course, in the present case as {well as} in (\ref{new1}), the non--conservation of energy-momentum tensor introduces an extra term in the TOV equation. Explicitly for an isotropic material content this equation reads  
\begin{equation}\label{new6}
\frac{{\nu^\prime}}{2}\,({\rho}+{p})+\frac{d{p}}{dr}-\frac{\lambda}{4\lambda-1}\frac{d}{dr}\left(\rho-3p\right)=0,    
\end{equation}
\begin{equation}\label{new7}
\frac{{\nu^\prime}}{2}\,({\rho}+{p})+\frac{d{p}}{dr}-2\beta p\frac{dR}{dr}=0.    
\end{equation}
It should be noted that the nature (attractive or repulsive) of the force introduced by the extra term in the above expressions in principle depends on the sign of the constants $\lambda$ and $\beta$ respectively. Nevertheless, these {parameters} must be constrained by solar system tests. 

In comparing the Newtonian limit reproduced by Rastall gravity theory and (\ref{new1}) for a perfect fluid matter distribution one has
\begin{equation}\label{new3}
\Delta\Phi-4\pi B\rho\Phi=4 A\pi\rho,
\end{equation}
where $\Phi$ is the Newtonian gravitational potential, $A\equiv (3\lambda-1)/(6\lambda-1)$ and $B\equiv -2\lambda/(6\lambda-1)$. As can be seen the Newtonian limit in the Rastall framework does not reproduce the well known Poisson's equation. In fact, due to the non-minimal coupling an extra term appears. This new term in Eq. (\ref{new3}), shows that in the classical limit the source depends on the potential $\Phi$ of gravitational field. This
puts a strong constraint on $\lambda$. It is worth mentioning that $\lambda$ should be zero in order to obtain the
usual Poisson equation for weak gravitational fields. Otherwise for large mass and strong gravitational field regime, expression (\ref{new3}) could play a crucial role in the cosmological scenario \cite{rawaf}. Moreover, if $\rho=\text{constant}$ Eq. (\ref{new3}) transforms into Sileeger equation \cite{hugo}. On the other hand, the Newtonian limit for the theory (\ref{new1}) provides the following gravitational potential $\Phi$,
\begin{equation}\label{new5}
\Phi=-\frac{1}{2}\left[\xi+\text{Ln}\left(1-\xi\right)\right],   
\end{equation}
where $\xi=\xi(r)$. In this case the gravitational potential does not coincide with Newtonian one, this is because an additional logarithmic term appears. So, in both cases the corresponding Newtonian limit {goes} beyond the classical one. This suggests that such
modifications could introduce new insights and implications.
Another {interesting} point to be compared here, are the cosmological and cluster dynamic consequences. In this respect, Rastall gravity theory has been theoretically tested as a feasible framework to explain the $\Lambda$CDM model issues. In this direction abandon exotic fluid such as Chaplygin gas to explain the existence of dark energy is a viable way. An
interesting way out is to use a non--canonical self--interacting scalar field as suggested
by Rastall gravity theory \cite{fabris} or analyzed cosmological models at the background as well as perturbation level \cite{dutta}. In this sense, Rastall's proposal has proven to be a good candidate to explain such problems. On the other hand the theory (\ref{new1}) is also a good alternative to explain dark components presence. Bertolami and P$\text{\'{a}}$ramos  \cite{bertolami3} studied and compared the known dark matter density profiles through an
appropriate power-law coupling $f_{2}=(R/R_{0})^{n}$ (with $n<0$), where was shown 
that dark matter dominates at cosmological scales. Although both {theories} share the same characteristic \i.e, a non--conservative energy--momentum tensor and also serve as an alternative theories of gravity which results are commensurable with those provided by GR, it is not possible to move from one formulation to another. The main fact is that Rastall's field equations are not obtained from a variational principle as they are in the case of the theory (\ref{new1}) \cite{bertolami}.

\section{Field equations: Multiple sources}\label{sec3}
In this section we describe the field equations for multiple matter sources. So, the standard field equations are given by 
\begin{equation}\label{eq10}
R_{\mu\nu}-\frac{1}{2}Rg_{\mu\nu}=\bar{T}^{(\text{tot})}_{\mu\nu},    
\end{equation}
where $\bar{T}^{(\text{tot})}_{\mu\nu}$ stands for
\begin{equation}\label{eq11}
\bar{T}^{(\text{tot})}_{\mu\nu}=T^{(\text{eff})}_{\mu\nu}+\alpha\theta_{\mu\nu}. 
\end{equation}
The new
sector $\theta_{\mu\nu}$ always can be seen as corrections to the theory
and be consolidated as part of an effective energy--momentum tensor. This extra source could represent a scalar, vector or tensor fields and {introduces} anisotropies within the self--gravitating systems. It is coupled to the matter sector through a dimensionless constant parameter \i.e,   $\alpha$. On the other hand, $T^{(\text{eff})}_{\mu\nu}$ represents the usual matter sector, that is isotropic, anisotropic, or charged distributions, among others. In the present case $T^{(\text{eff})}_{\mu\nu}$ is given by Eqs. (\ref{eq8})-(\ref{eq9}). As we are interested in studying spherically symmetric and static fluid spheres, next we regard the most general line element in the standard Schwarzschild like coordinates  $\{t,r,\phi,\theta\}$ to be
\begin{equation}\label{eq12}
{ds}^{2}={{e}^{\nu(r)}}{{dt}^{2}}-{{e}^{\eta(r)}}{{dr}^{2}}-{r}^{2}({{d\theta}^{2}}+{{sin}^{2}}\theta{{d\phi}^{2}}).   
\end{equation}
The staticity of this space-time is ensured by considering $\nu$ and $\eta$ as functions of the radial coordinate $r$ only. Putting together equations (\ref{eq8}), (\ref{eq9}), (\ref{eq10}) and (\ref{eq12}) one arrives at the following set of equations
\begin{eqnarray}\label{eq13}
{{\rm e}^{-\eta}}\left( {\frac {\eta^{{\prime}}}{r}}-\frac{1}{r^2}\right)+\frac{1}{r^2}&=&{\rho}^{(\text{eff})}+\alpha\theta^{t}_{t},~~~~~~~~ \\ \label{eq14}
{{\rm e}^{-\eta}} \left( {\frac {\nu^{{\prime}}}{r}}+\frac{1}{r^2}\right) -\frac{1}{r^2}&=&{p}^{(\text{eff})}-\alpha\theta^{r}_{r}, ~~~~~~~~\\ \label{eq15}
\frac{{\rm e}^{-\eta}}{4} \left(2 \nu^{{\prime\prime}}+\nu^{\prime 2}+2{\frac {\nu^{{\prime}}-\eta^{{\prime}}}{r}}-\nu^{\prime}\eta^{\prime} \right) &=&{p}^{(\text{eff})}-\alpha\theta^{\varphi}_{\varphi}.~~~~~~~
\end{eqnarray}
where
\begin{eqnarray}\label{eq16}
{{\rho}}^{(\text{eff})}&=&\frac{\left(3\lambda-1\right)\rho+3\lambda p}{4\lambda-1}, \\ \label{eq17}
-{p}^{(\text{eff})}&=&-\frac{\left(\lambda-1\right)p+\lambda \rho}{4\lambda-1}.
\end{eqnarray}
The corresponding conservation law $\nabla^{\mu}\bar{T}^{\textrm{(tot)}}_{\mu\nu}=0$ associated with the system (\ref{eq13})-(\ref{eq15}) reads
\begin{equation}\label{eq18}
\begin{split}
-\frac{d{p}^{(\text{eff})}}{dr}-\alpha\left[ \frac{\nu^{\prime}}{2}\left(\theta^t_t-\theta^r_r\right)-\frac{d \theta^r_r}{dr}+\frac{2}{r}\,(\theta^\varphi_\varphi-\theta^r_r)\right]& \\-\frac{\nu^\prime}{2} ({\rho}^{(\text{eff})}+{p}^{(\text{eff})})=0.
\end{split}
\end{equation}
It is found that the system of non--linear differential equations (\ref{eq13})--(\ref{eq15})
consists of seven unknown functions, the metric potentials $\{\eta, \nu\}$, the thermodynamic observables $\{\rho^{(\text{eff})},
p^{(\text{eff})}\}$ and the components of the extra source $\{\theta^{t}_{t}, \theta^{r}_{r}, \theta^{\varphi}_{\varphi}\}$ . In order to find these {unknowns} we adopt a 
systematic approach. Furthermore, for the
system (\ref{eq13})--(\ref{eq15}), the matter content (total energy--density, total radial
pressure and total tangential pressure) can be identified as
\begin{eqnarray}\label{eq19}
\bar{\rho}^{(\text{tot})}(r)&=&\rho^{(\text{eff})}(r)+\alpha\theta^{t}_{t}(r) \\ \label{eq20}
\bar{p}^{(\text{tot})}_{r}(r)&=&p^{(\text{eff})}(r)-\alpha\theta^{r}_{r}(r) \\ \label{eq21}
\bar{p}^{(\text{tot})}_{t}(r)&=&p^{(\text{eff})}(r)-\alpha\theta^{\varphi}_{\varphi}(r).
\end{eqnarray}
It is clear that an anisotropic behaviour arises into the system due to the {presence} of the $\theta$--sector if $\theta^{r}_{r}\neq\theta^{\varphi}_{\varphi}$. So, in order to measure the anisotropic behaviour we {define} the anisotropy factor as follows
\begin{equation}\label{eq22}
\Delta=\bar{p}^{(\text{tot})}_{t}-\bar{p}^{(\text{tot})}_{r}=\alpha\left(\theta^{r}_{r}-\theta^{\varphi}_{\varphi}\right).
\end{equation}
At this stage the system of Eqs. (\ref{eq13})--
(\ref{eq15}) could be treated as an anisotropic fluid, with five unknown functions,
namely, the two metric functions $\nu$ and $\eta$, and the total
functions in Eqs. (\ref{eq19})-(\ref{eq21}). However, we are going
to implement a different approach, as explained below.

\section{Gravitational Decoupling: A MGD Approach}\label{sec4}
As said before, gravitational decoupling by MGD scheme becomes a simple and powerful tool to extent spherically and static isotropic fluid solutions to anisotropic domains \cite{r57}. To see how this approach works let us start by turning off the coupling $\alpha$, so we are describing a perfect fluid solution given by $\{\xi,\mu,\rho^{(\text{eff})},p^{(\text{eff})}\}$, being $\xi$ and $\mu$ the corresponding metric functions. The metric (\ref{eq12}) now reads 
\begin{equation}\label{eq23}
{ds}^{2}={{e}^{\xi(r)}}{{dt}^{2}}-\frac{{dr}^{2}}{\mu(r)}-{r}^{2}({{d\theta}^{2}}+{{sin}^{2}}\theta{{d\phi}^{2}}),
\end{equation}
where $\mu(r)=1-\frac{2m_{GR}}{r}$  
is the standard GR expression containing the mass function of the fluid configuration. Next, to see the effects of the $\theta$--sector on the perfect fluid distribution we turn on the coupling $\alpha$. These effects can be encoded in the geometric deformation
undergone by the perfect fluid geometry $\{\xi,\mu\}$ in Eq. (\ref{eq23}), namely
\begin{eqnarray}\label{eq24}
\xi\rightarrow \nu&=&\xi+\alpha h \\ \label{eq25}
\mu\rightarrow e^{-\eta}&=&\mu+\alpha f,
\end{eqnarray}
where $h$ and $f$ are the deformations introduced in the temporal
and radial metric {components}, respectively. {It is} worth mentioning that the foregoing deformations are purely radial functions, this feature ensures the spherical symmetry of the solution. The MGD scheme consists in set off either $h$ or $f$. In this opportunity we set $h=0$, it means that the temporal component remains unchanged and the anisotropy lies on the radial component \cite{r57}. So, we have
\begin{equation}\label{eq26}
\mu(r)\rightarrow e^{-\eta(r)}=\mu(r)+\alpha f(r).    
\end{equation}
Upon replacing Eq. (\ref{eq26}) in the  equations (\ref{eq13})-(\ref{eq15}), the system splits into two sets of equations. The first one corresponds to
$\alpha= 0$ that is, perfect fluid matter distribution

\begin{eqnarray}\label{eq27}
-\frac{\mu^{\prime}}{r}-\frac{\mu}{r{^2}}+\frac{1}{r^{2}}&=&  {\rho}^{(\text{eff})} ,~~~~~~~~~\\ \label{eq28}
 \mu \left( {\frac {{\nu}^{{\prime}}}{r}}+\frac{1}{r^2}\right) -\frac{1}{r^2}&=&  {p}^{(\text{eff})},~~~~~~~~~ \\ \label{eq29}
 \frac{\mu}{4} \left( {2\nu}^{{\prime\prime}}+{\nu}^{{\prime\,2}}+2\frac{{\nu}^{{\prime}}}{r}\right)+\frac{\mu^{\prime}}{4}\,\left( {\nu}^{{\prime}}+\frac{2}{r}  \right) &=& {p}^{(\text{eff})}.~~~~~~~~~
\end{eqnarray}
From now on we shall call the above system of equations the Einstein-Rastall system. It can be solved for $\rho$ and $p$ by using Eqs.(\ref{eq16}) and (\ref{eq17}), in order to express these quantities as functions of the metric potentials only \cite{r45}. Explicitly we have
\begin{equation}\label{eq30}
-\frac{\mu^{\prime}}{r}-\frac{\mu}{r{^2}}+\frac{1}{r^{2}} -\lambda\left[-\mu\left(\frac{4}{r^{2}}+\frac{3\nu^{\prime}}{r}\right)+\frac{4}{r^{2}}-\frac{\mu^{\prime}}{r}\right]=\rho,
\end{equation}
\begin{eqnarray}
\label{eq31}
 \mu \left( {\frac {{\nu}^{{\prime}}}{r}}+\frac{1}{r^2}\right) -\frac{1}{r^2}+\lambda\left[\frac{4}{r^{2}}-\mu\left(\frac{4}{r^{2}}+\frac{3\nu^{\prime}}{r}\right)-\frac{\mu^{\prime}}{r}\right]=p,
\end{eqnarray}

\begin{equation}\label{neweq}
\begin{split}
 \frac{1}{4}\left[\mu\left(2\nu^{\prime\prime}+\nu^{\prime 2}+2\frac{\nu^{\prime}}{r}\right)+\mu^{\prime}\left(\nu^{\prime}+\frac{2}{r}\right)\right]
&\\+\lambda\left[-\mu\left(\frac{4}{r^{2}}+\frac{3\nu^{\prime}}{r}\right)+\frac{4}{r^{2}}-\frac{\mu^{\prime}}{r}\right]=p.
\end{split}    
\end{equation}
As was pointed out earlier, both $\rho$ and $p$ after some algebraic manipulations, in their own expressions contain the Rastall information as was expected. Besides by putting $\lambda=0$ in Eqs. (\ref{eq30})-(\ref{neweq}) one recovers the original GR field equations for isotropic matter distributions. Furthermore, by adding (\ref{eq30}) to (\ref{eq31}) one regains the usual inertial mass density $\rho+p$ which is given by
\begin{equation}\label{eq32}
\rho+p=\frac{\mu \nu^{\prime}-\mu^{\prime}}{r}.
\end{equation}
Another interesting point to be noted here, is that the isotropic condition is exactly the same like in GR \i.e,
\begin{equation}\label{eq33}
\begin{split}
4\left(1-\mu\right)+2r\left(\mu^{\prime}-\mu\nu^{\prime}\right)+r^{2}\left(2\mu\nu^{\prime\prime}+\mu\nu^{\prime 2}+\mu^{\prime}\nu^{\prime}\right)=0.   
\end{split}
\end{equation}
Equation (\ref{eq33}) says that any solution describing a perfect fluid matter distribution in GR is also a solution in the arena of Rastall gravity theory. Obviously there is a subtlety, since both GR and Rastall theory share only the geometrical content but not the material one, is in this sense that "any" solution to Einstein theory of gravity can be seen as a solution in the gravitational Rastall framework.  
So, the other set of equations corresponding to the factor $\theta$ are given by,

\begin{eqnarray}\label{eq34}
 -\frac{f{^\prime}}{r}-\frac{f}{r^2}&=&  \theta^{t}_{t} ,~~~~~~~~~ \\ \label{eq35}
-f \left( {\frac {{\nu}^{{\prime}}}{r}}+\frac{1}{r^2}\right) &=&  \theta^{r}_{r} ~~~~~~~~~ \\ \label{eq36}
-\frac{f}{4} \left( {2\nu}^{{\prime\prime}}+{\nu}^{{\prime\,2}}+2\frac{{\nu}^{{\prime}}}{r}\right)-\frac{f^{\prime}}{4}\left( {\nu}^{{\prime}}+\frac{2}{r}  \right)&=& \theta^{\varphi}_{\varphi}.~~~~~~~~~
\end{eqnarray}

The {sets} of equations (\ref{eq30})-(\ref{neweq}) and (\ref{eq34})-(\ref{eq36}) {satisfy} the following conservation equations,

\begin{eqnarray}\label{eq37}
\frac{{\nu^\prime}}{2}\,({\rho}+{p})+\frac{d{p}}{dr}-\frac{\lambda}{4\lambda-1}\frac{d}{dr}\left(\rho-3p\right)=0, \\ \label{eq38}
-\frac{{\nu^\prime}}{2}\,(\theta^t_t-\theta^r_r)+\frac{d \theta^r_r}{dr}-\frac{2}{r}\,(\theta^\varphi_\varphi-\theta^r_r)=0 
\end{eqnarray}
 
We note that the linear combination of conservation equations (\ref{eq37}) and (\ref{eq38}) via. coupling constant $\alpha$ provides the conservation equation for the energy momentum tensor $\bar{T}^{\mu \textrm{(tot)}}_{\nu}={T}^{\mu{(\text{eff})}}_{\nu} + \alpha\theta^{\mu}_{\nu}$,
as follows 
\begin{eqnarray}
\label{eq39}
-\frac{d{p}}{dr}-\alpha\left[ \frac{{\nu^\prime}}{2}\,(\theta^t_t-\theta^r_r)-\frac{d \theta^r_r}{dr}+\frac{2}{r}\,(\theta^\varphi_\varphi-\theta^r_r)\right]-\frac{{\nu^\prime}}{2}\,({\rho}+{p})\nonumber\\+\frac{\lambda}{4\lambda-1}\frac{d}{dr}\left(\rho-3p\right)=0.
\end{eqnarray}
The Eq. (\ref{eq39}) is the same expression as Eq. (\ref{eq18}) but in an explicit form. Furthermore, as can be seen there is an extra term (the last one) in (\ref{eq39}), the so called Rastall force (or {simply} the Rastall contribution). This additional term could in principle be attractive or repulsive in nature, due to its behaviour depends on the sign of the Rastall parameter $\lambda$.

At this point it is necessary to comment that from now on the total energy-momentum tensor $T^{\text(tot)}_{\mu\nu}$ will be defined by the following components
\begin{eqnarray}\label{41}
\rho^{\textrm(tot)}(r)&=&\rho(r)+\alpha\theta^{t}_{t}(r), \\ \label{42}
p_{r}^{\textrm(tot)}(r)&=&p(r)-\alpha\theta^{r}_{r}(r), \\ \label{43}
p_{t}^{\textrm(tot)}(r)&=&p(r)-\alpha\theta^{\varphi}_{\varphi}(r),
\end{eqnarray}
where $\rho$ and $p$ are given by Eqs. (\ref{eq30}) and (\ref{eq31}), respectively. This equations contain the additional geometric terms provided by the Rastall contribution. In this way, as we will see in the following sections, there will be a full affect of the Rastall sector in the decoupler function $f(r)$ and consequently in the $\theta$-sector, as expected. Besides, the redefinition (\ref{41})-(\ref{43}) does not change the anisotropy factor $\Delta$ definition given by Eq. (\ref{eq22}).

\section{Exterior space-time: Junction conditions}\label{sec5}
A crucial aspect in the study of stellar distributions is the junction conditions. These provide smooth matching of the interior $\mathcal{M^{-}}$ and the exterior $\mathcal{M^{+}}$
geometries at the surface $\Sigma$ (defined by $r=R$) of the stellar object, to investigate some significant characteristics of {its} evolution. To study how the junction conditions work in this context we will assume that the inner stellar
geometry $\mathcal{M^{-}}$ is given by the MGD metric,
\begin{equation}\label{eq40}
ds^{2}={{e}^{\nu(r)}}{{dt}^{2}}-\left(1-2\frac{m(r)}{r}\right)^{-1}{{dr}^{2}}-{r}^{2}({{d\theta}^{2}}+{{sin}^{2}}\theta{{d\phi}^{2}}),
\end{equation}

where the interior mass function in this case is given by
\begin{equation}\label{eq41}
4\pi\int^{r}_{0}\rho^{(\text{tot})} r^{2}dr\equiv m(r)=m_{GR}(r)+m_{\lambda}(r)-\alpha\frac{r}{2}f(r),
\end{equation}
where we have defined $m_{GR}(r)$ as
\begin{equation}\label{eq522}
m_{GR}(r)=\frac{r}{2}\left[1-\mu(r)\right], 
\end{equation}
where from now on we shall call the total mass coming from the GR sector as $m_{GR}(R)=M_{0}$. On the other hand, the $m_{\lambda}(r)$ term is equal to
\begin{equation}\label{mlambda}
m_{\lambda}(r)=\frac{\lambda}{2}\int^{r}_{0}r^{2}\left[-\mu\left(\frac{4}{r^{2}}+\frac{3\nu^{\prime}}{r}\right)+\frac{4}{r^{2}}-\frac{\mu^{\prime}}{r}\right]\,dr,  
\end{equation}
where at the boundary becomes $M_{\lambda}=m_{\lambda}(R)$, the Rastall mass hereinafter. So, when $\alpha=\lambda=0$ the familiar gravitational mass definition is recovered. At this stage it is of interest to contrast the so called binding energy in this context with what is reported in GR. In few words the binding energy (B.E.) is the difference between the total mass and the proper mass. Explicitly it reads
\begin{equation}\label{BE}
\text{B.E.}=m(R)-m_{p}(R), 
\end{equation}
where the proper mass $m_{p}$ is given by
\begin{equation}\label{proper}
m_{p}(R)=4\pi\int^{R}_{0}\frac{r^{2}\rho}{\sqrt{1-\frac{2{m}}{r}}}dr.  
\end{equation}
Since the factor $\sqrt{1-\frac{2m}{r}}$, appearing in the proper mass $m_{p}$ is less than the unity. Then the proper mass is greater than the total mass, hence $B.E.<0$. Respect to the GR case, in the present one the proper mass will be little bit different. The main difference is introduced in the factor $m/r$. As it is well known in the isotropic (uncharged) case this factor at the boundary $\Sigma$ is bounded by the Buchdhal limit \cite{r105} \i.e, $\frac{M_{0}}{R}\leq \frac{4}{9}$. On the other hand in the anisotropic (uncharged) case, the above limit can be overcome \cite{r107}. In this opportunity the ratio $m/r$ is altered by the Rastall and MGD contributions. So, we have
\begin{equation}
\frac{m}{r}=\frac{m_{GR}+m_{\lambda}}{r}-\frac{\alpha}{2}f(r).    
\end{equation}
Then,
\begin{equation}\label{buch}
1-2\frac{\left(m_{GR}+m_{\lambda}\right)}{r}>\alpha f(r).    
\end{equation}
It is obvious that the above constraint impose some restriction in order to avoid non-physical situations. Besides, (\ref{buch}) also imposes some restrictions on the constant $\alpha$, since $\alpha$ is not restricted to be a strictly positive quantity. So, the proper mass in this case may be greater or less than that the GR case. Therefore the B.E. will be change according the MGD contribution.    

Next, the internal manifold (\ref{eq40}) should be joined in a smoothly way with outer space-time. This exterior manifold in principle could {contain} some contributions coming from the $\theta$-sector. So, {this} means that the exterior space-time surrounding the compact structure is no longer a vacuum space-time. The most general outer manifold is described by
\begin{equation}\label{eq42}
ds^{2}={{e}^{\nu^{+}(r)}}{{dt}^{2}}-e^{\eta^{+}(r)}{{dr}^{2}}-{r}^{2}({{d\theta}^{2}}+{{sin}^{2}}\theta{{d\phi}^{2}}).
\end{equation}
To match the internal configuration (\ref{eq40}) with the exterior one (\ref{eq42}) we employ the well known Israel--Darmois (ID hereinafter) junction conditions \cite{r95,r96} (for a recent and more clear discussion of how these conditions work see \cite{r99,r100,r101}). These conditions are the most general ones. The ID matching conditions involve the first and second fundamental forms. The first fundamental form express the continuity of the metric potentials across the boundary $\Sigma$. More specifically, the metric potentials describe the intrinsic geometry of the manifolds. So, the continuity of the first fundamental form across the boundary of the compact structure, reads
\begin{equation}\label{eq43}
\left[ds^{2}\right]_{\Sigma}=0,
\end{equation}
concisely
\begin{equation}\label{eq44}
e^{\nu^{-}(r)}|_{r=R^{-}}=e^{\nu^{+}(r)}|_{r=R^{+}},
\end{equation}
and
\begin{equation}\label{eq45}
1-\frac{2M}{R^{-}}=e^{-\eta^{+}(r)}|_{r=R^{+}},
\end{equation}
being $M=m(R)$ the total gravitational mass contained by the fluid sphere. The second fundamental form is related with the continuity of the extrinsic curvature $K_{\mu\nu}$ induced by $\mathcal{M}^{-}$ and $\mathcal{M}^{+}$ on $\Sigma$. The continuity of $K_{rr}$ component across $\Sigma$ yields to
\begin{equation}\label{eq46}
\left[p^{(\text{tot})}_{r}(r)\right]_{\Sigma}=\left[p(r)-\alpha\theta^{r}_{r}(r)\right]_{\Sigma}=0.
\end{equation}
At this stage some comments are in order. First, $p^{(\text{tot})}_{r}$ has a little different fashion in comparison with the expression (\ref{eq20}) since $\rho$ and $p$ were separated implying that the Rastall terms are no longer contained in $p^{(\text{eff})}$  through $\lambda$ as shown Eq. (\ref{eq9}). Now the terms coming from the Rastall sector are encoded in separate expressions for $\rho$ and $p$ given by Eqs. (\ref{eq30})--(\ref{eq31}). From this point of view it is clear how Rastall contribution comes into the field equations. Moreover, from now on we shall denote the Rastall input as follows 
\begin{equation}\label{eq47}
F_{\lambda}(r)=\lambda\left[-\mu\left(\frac{4}{r^{2}}+\frac{3\nu^{\prime}}{r}\right)+\frac{4}{r^{2}}-\frac{\mu^{\prime}}{r}\right].
\end{equation}
It should be noted that the form of $F_{\lambda}$ depends on the choice of $T_{\mu\nu}$ which in our case is given by Eq. (\ref{eq7}) describing a perfect fluid matter distribution.
Hence, $p(r)$ in Eq. (\ref{eq46}) is given by Eq. (\ref{eq31}). 
Second, in this way the Rastall sector will come into the $\theta$--sector through the decoupler function $f(r)$ (as we will see later) in order to see the effects on it. So Eq. (\ref{eq46}) reads
\begin{equation}\label{eq48}
\left[p(r)-\alpha\theta^{r}_{r}(r)\right]_{r=R^{-}}=
\left[-F_{\lambda}(r)-\alpha\theta^{r}_{r}(r)\right]_{r=R^{+}}.
\end{equation}
Equation (\ref{eq48}) tells us that the outer space-time receives contributions from the $\theta$--sector, as well as from the Rastall non--minimal coupling matter assumption. In this respect, in the study of compact structures within the framework of modified gravity theories such as $f(R)$, the exterior space--time receives contributions from the inclusion of higher order derivative terms coming from the Ricci scalar. In principle, these contributions can alter or introduce some modifications on the usual junction conditions. Moreover, the outer manifold could be different from the usual ones \i.e Schwarzschild vacuum solution, Reissner--Nordstr\"{o}m, for example. At this stage and based on the previous discussion, a couple of comments are pertinent in order to clarify how to proceed in modified type gravity theories. In this direction, Capozziello et.al \cite{r13} have argued that in the $f(R)$ domain the mass--radius profile undergoes modifications due to the presence of high order curvature terms such as $R^{2}$, $R^{3}$, etc. Besides, in \cite{r98} was discussed the well--known ID matching conditions in the framework of $f(R)$ gravity in considering both isotropic and anisotropic matter distributions.  They conclude that ID matching conditions are not satisfied at all in the $f(R)$ gravity arena. 
However, in the present situation, one could dropped out the Rastall contribution $F^{+}_{\lambda}$ from the external space--time. To do so, one needs to consider an outer geometry free from material content \i.e, a vanishing energy-momentum tensor $T^{+}_{\mu\nu}=0$. Then from Eqs. (\ref{eq4}) and (\ref{eq5}) one arrives to
\begin{equation}\label{eq49}
G_{\mu\nu}=0.
\end{equation}
The above expression implies (as said before) that both Einstein and Rastall gravity theories share exactly the same vacuum solution \i.e, the exterior Schwarzschild solution. If one wants to consider contributions coming from the Rastall sector, the outer space--time is no longer vacuum, since it is filled by an effective cosmological constant describing a (anti) de Sitter space--time \cite{r50}. So, Eq.(\ref{eq48}) becomes 
\begin{equation}\label{eq50}
\left[p(r)-\alpha\theta^{r}_{r}(r)\right]_{r=R^{-}}=
\left[-\alpha\theta^{r}_{r}(r)\right]_{r=R^{+}}.
\end{equation}
It remains to be analyzed how the $\theta$--sector affects the exterior geometry. In this case the external solution comes from solving the field equations
\begin{equation}\label{eq51}
R_{\mu\nu}-\frac{R}{2}g_{\mu \nu}=\alpha\theta_{\mu\nu},  
\end{equation}
in conjunction with (\ref{eq42}). Hence, the resulting outward geometry is described by
\begin{eqnarray}
\label{eq52}
ds^{2}=\left(1-\frac{2{M_{\text{Sch}}}}{r}\right)dt^{2}- \left(1-\frac{2{M_{\text{Sch}}}}{r}+\alpha g(r)\right)^{-1}dr^{2}\nonumber\\ -r^{2}d\Omega^{2},
\end{eqnarray} 
where $g(r)$ is the  geometric deformation of the exterior Sch-warzschild space--time associated to the source $\theta_{\mu\nu}$, and ${M_{\text{Sch}}}$ {denotes} the Schwarzschild mass. By using Eq. (\ref{eq35}) in (\ref{eq50}) we obtain 
\begin{equation}\label{eq53}
\begin{split}
p(R)+\alpha f(R)\left(\frac{1}{R^{2}}+\frac{\nu^{\prime}(R)}{R}\right)=\alpha g(R)
\bigg[\frac{1}{R^{2}}+&\\
\frac{2{M_{\text{Sch}}}}{R^{3}}\frac{1}{\left(1-\frac{2{M_{\text{Sch}}}}{R}\right)}\bigg], 
\end{split}
\end{equation}
where $R^{-}=R^{+}=R$ at the surface. It should be noted that if the geometric deformation function $g(r)$ of the outer manifold is taken to be null, then one recovers the original Schwarzschild exterior solution. In consequence Eq. (\ref{eq53}) leads to the condition
\begin{equation}\label{eq54}
\begin{split}
{p}_{r}^{(\text{tot})}(R)=p(R)+\alpha f(R)\left(\frac{1}{R^{2}}+\frac{\nu^{\prime}(R)}{R}\right)=0.
\end{split}
\end{equation}
Equation (\ref{eq54}) is an important result, since the compact object will be in equilibrium in a true exterior space--time without material content (vacuum) only if the total radial pressure at the surface vanishes.
The condition (\ref{eq54}) determines the size of the object \i.e the radius $R$, which means that the material content is confined within the region $0\leq r\leq R$. Furthermore the continuity of the remaining components of the extrinsic curvature $K_{\theta\theta}$ and $K_{\phi\phi}$ yield to
\begin{equation}\label{eq55}
m(R)=M.
\end{equation}

\section{Stellar interior: Tolman IV model}\label{sec6}
In this section we solve the set of equations (\ref{eq34})--(\ref{eq36}) by imposing some suitable constraints on the $\theta_{\mu\nu}$ components in order to obtain the deformation function $f(r)$ and then compute the full energy--momentum tensor $T^{(\text{tot})}_{\mu\nu}$. Among all the possibilities, to tackle the system of equations  (\ref{eq34})--(\ref{eq36}) we follow the same procedure as given in \cite{r57}. The imposition of some extra information is necessary in order to close the system of equations (\ref{eq34})--(\ref{eq36}). Furthermore, to obtain the deformation function $f(r)$ also is necessary provide a seed solution satisfying equations (\ref{eq30})--(\ref{neweq}). To illustrate how gravitational decoupling by means of MGD works in the Rastall gravity scenario, we apply it to the well known Tolman IV solution. This space--time was already studied in the context of MDG in \cite{r57} and in the framework of Rastall theory \cite{r45}. So, {this} allows us to compare the resulting deformed solution immersed in an anisotropic scenario with previous results already obtained and therefore establish whether the study of the compact structures whitin the arena of Rastall gravity $+$ gravitational decoupling by MGD approach is plausible, when the matter distribution contains local anisotropies. Before to proceed we present the well--known Tolman IV space--time within the Rastall framework, which is described by the following metric potentials  
\begin{eqnarray}\label{eq56}
e^{\nu(r)}&=&B^{2}\left(1+\frac{r^{2}}{A^{2}}\right), \\ \label{eq57}
\mu(r)&=&\frac{\left(1-\frac{r^{2}}{C^{2}}\right)\left(1+\frac{r^{2}}{A^{2}}\right)}{\left(1+2\frac{r^{2}}{A^{2}}\right)},
\end{eqnarray}
and characterized by the following thermodynamic observables (in the Rastall context)
\begin{equation}\label{eq58}
\begin{split}
\rho=\frac{1}{C^{2}\left(A^{2}+2r^{2}\right)^{2}}\bigg\{2\lambda\bigg[r^{2}\bigg(2C^{2}-11A^{2}\bigg)-3\bigg(A^{4}&\\+4r^{4}\bigg)\bigg]+3A^{4}+A^{2}\bigg(3C^{2}+7r^{2}\bigg)+2r^{2}\bigg(C^{2}+3r^{2}\bigg)\bigg\},
\end{split}
\end{equation}
\begin{equation}\label{eq59}
\begin{split}
p=\frac{1}{C^{2}\left(A^{2}+2r^{2}\right)^{2}}\bigg\{2\lambda\bigg[r^{2}\bigg(11A^{2}-2C^{2}\bigg)+3\bigg(A^{4}&\\+4r^{4}\bigg)\bigg]+\bigg(A^{2}+2r^{2}\bigg)\bigg(C^{2}-A^{2}-3r^{2}\bigg)\bigg\}.
\end{split}
\end{equation}
At this point a couple of comments are in order. First, it is worth mentioning that taking the limit $\lambda\rightarrow 0$ expressions (\ref{eq58})--(\ref{eq59}) are the corresponding energy--density and isotropic pressure of the original Tolman IV solution satisfying Einstein field equations. Second, as was pointed out earlier expressions (\ref{eq56})--(\ref{eq59}) confirm that the geometrical description of any perfect fluid solution of Einstein gravity theory is also a solution from the geometrical point of view in the Rastall gravity scenario, but containing a more complicated thermodynamical behaviour due to the non--minimal coupling matter introduced by the Rastall sector. Now, we proceed to close the problem by using the mimic constraint scheme.

\subsection{$\theta$-effects: Mimicking the pressure for anisotropy} \label{A}

\begin{figure*}
\centering
\includegraphics[width=0.42\textwidth]{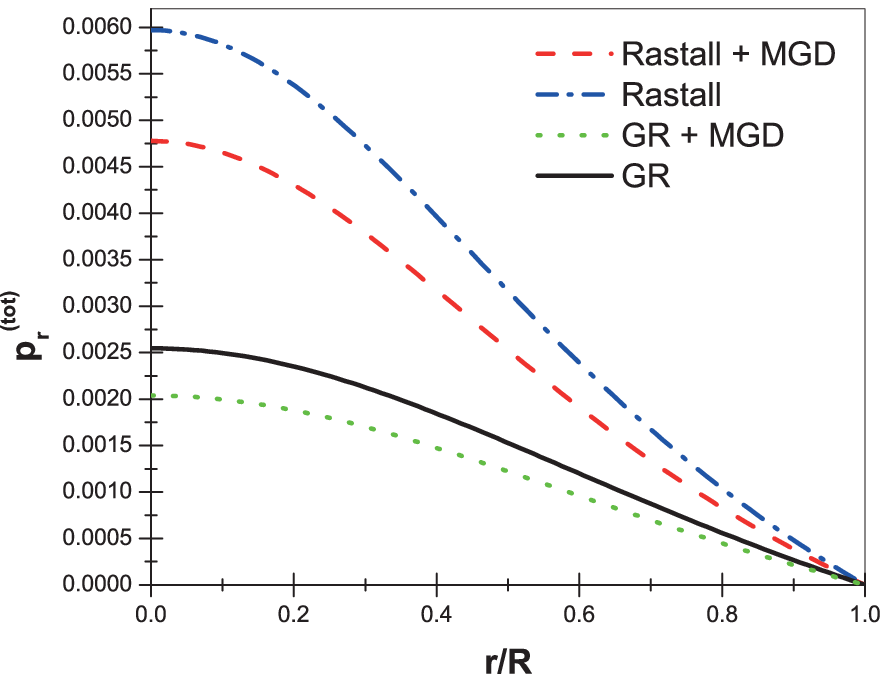}     ~~~
\includegraphics[width=0.42\textwidth]{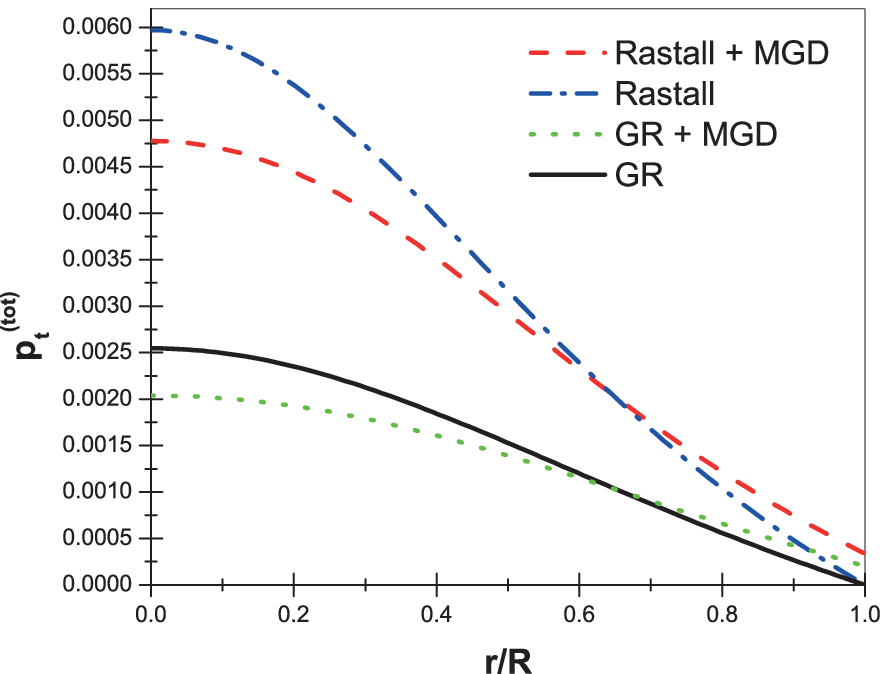}      \\
\includegraphics[width=0.42\textwidth]{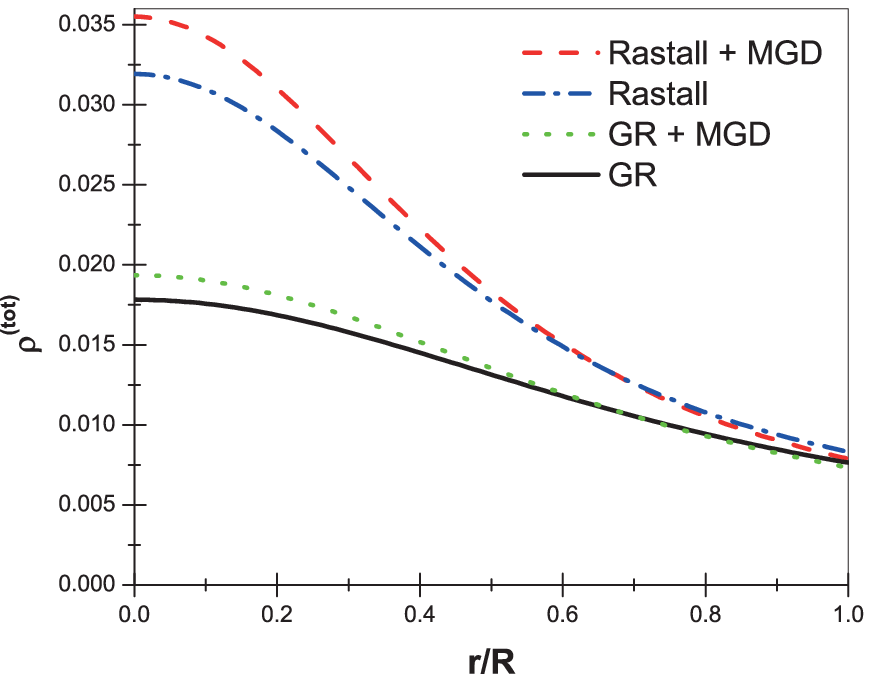}      ~~~
\includegraphics[width=0.42\textwidth]{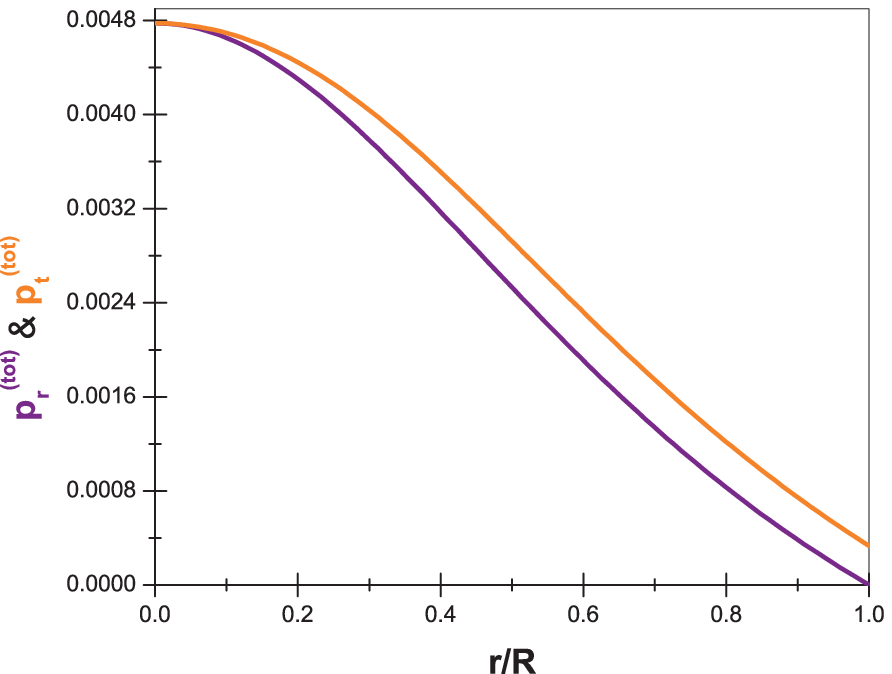}      
\caption{\textbf{Mimic Constraint} $p(r)=\theta^{r}_{r}$. To obtain the trend of the principal thermodynamic observables we have considered throughout the study the following mass--radius ratio $M_{0}/R=0.2$. Moreover, the red curve (dashed) representing Rastall + MGD corresponds to $\alpha=0.2$ and $\lambda=-0.09$, for  the blue (dashed--dotted) one  corresponding to pure Rastall gravity, $\alpha$ and $\lambda$ are $0.0$ and $-0.09$, respectively. Next, the green (short--dashed) line corresponding to GR + MGD takes $\alpha=0.2$ and $\lambda=0.0$. Finally, the black curve (solid) is representing GR theory $\alpha=\lambda=0.0$. Upper row: 
{\bf Left panel} illustrates the monotonic behaviour from the center to the boundary of the total radial pressure at all points inside the structure. As it is observed this quantity vanishes at surface. The  {\bf Right panel}: shows the trend of the total tangential pressure inside the compact object. Lower row: {\bf Left panel} exhibits the behaviour of the total energy--density. Finally, the {\bf Right panel} displays a comparison between the total radial and total tangential pressure. It should be noted that the presence of anisotropies {causes} the pressures values to drift apart.}
\label{fig1}
\end{figure*}

The so called mimic constraints are some restrictions imposed at the level of the field equations (\ref{eq30})--(\ref{neweq}) and (\ref{eq34})--(\ref{eq36}) after introduce the decoupler mechanism (\ref{eq26}). In principle, these choices lead to well--behaved solutions, that is, free of undesired physical and mathematical behaviors such as singularities, non--decreasing thermodynamic functions, violation of causality condition, among others. However, other options can be considered, for example a direct and adequate representation for the geometric deformation function $f(r)$ \cite{r69,r70,r91} which satisfies the basic requirements of physical and mathematical admissibility, or relate only the $\theta$--sector components through a barotropic, polytropic or linear equation of state. In this opportunity, an acceptable interior solution is deduced when forcing the associated radial pressure $\theta^{r}_{r}$ to mimic the isotropic pressure $p(r)$. Explicitly it reads
\begin{equation}\label{eq60}
\theta^{r}_{r}(r)=p(r). 
\end{equation}
This constraint implies that the stress--energy tensor for the seed solution coincides with the anisotropy in the radial direction. Consequently Eq. (\ref{eq31}) and Eq. (\ref{eq35}) are {equals}. Thus, this yields to an algebraic general expression for the deformation function
\begin{equation}\label{eq61}
f(r)=-\mu(r)+\left[\frac{1}{r^{2}}-F_{\lambda}(r)\right]\left[\frac{\nu^{\prime}}{r}+\frac{1}{r^{2}}\right]^{-1},   
\end{equation}
where $F_{\lambda}(r)$ and $\mu(r)$ are given by Eq. (\ref{eq47}) and Eq. (\ref{eq57}), respectively and $\nu^{\prime}$ can be obtained from Eq. (\ref{eq56}). Then, the general minimally deformed radial metric potential is expressed as
\begin{equation}\label{lamb}
e^{-\eta}=\left(1-\alpha\right)\mu(r)+\alpha \left[\frac{1}{r^{2}}-F_{\lambda}(r)\right]\left[\frac{\nu^{\prime}}{r}+\frac{1}{r^{2}}\right]^{-1}.    
\end{equation}
The resulting expression of $f(r)$ after inserting the corresponding elements in Eq. (\ref{eq61}), is given by
\begin{eqnarray}
\label{eq62}
f(r)= \frac{-r^{2}}{C^{2}\left(A^{2}+2r^{2}\right)^{2}\left(A^{2}+3r^{2}\right)}\bigg[\bigg(A^{2}+r^{2}\bigg)\bigg(6A^{4}\lambda
\nonumber\\+22A^{2}r^{2}\lambda-4C^{2}r^{2}\lambda+24r^{4}\lambda\nonumber\\
-5A^{2}r^{2}+2C^{2}r^{2}-6r^{4}-A^{4}+A^{2}C^{2}\bigg)\bigg].  
\end{eqnarray}
In general, the deformed Tolman IV solution by virtue of (\ref{eq61}) can be expressed as 
\begin{eqnarray}\label{eq63}
ds^{2}=B^{2}\left(1+\frac{r^{2}}{A^{2}}\right)dt^{2}-\bigg[\left(1-\alpha\right) \frac{\left(1-\frac{r^{2}}{C^{2}}\right)\left(1+\frac{r^{2}}{A^{2}}\right)}{\left(1+2\frac{r^{2}}{A^{2}}\right)}\nonumber\\-\alpha\left[\frac{1}{r^{2}}-F_{\lambda}(r)\right]\left[\frac{\nu^{\prime}}{r}+\frac{1}{r^{2}}\right]^{-1} \bigg]dr^{2}-r^{2}d\Omega^{2}. 
\end{eqnarray}
Next, following the discussion in section \ref{sec5}, the constant parameters, namely $A$, $B$ and $C$ defining the interior solution can be obtained from 
\begin{equation}\label{eq64}
e^{\nu(r)}|_{r=R^{-}}=\left[\mu(r)+\alpha f(r)\right]|_{r=R^{-}}=1-2\frac{{M_{\text{Sch}}}}{R^{+}},
\end{equation}
where the Schwarzschild mass $M_{\text{Sch}}$ coincides at the boundary $\Sigma$ with total mass $M$ contained {by} the {sphere}. Furthermore, from Eqs. (\ref{eq46}) and (\ref{eq60}) we have 

\begin{equation}\label{eq65}
\left(1-\alpha\right)p(r)|_{r=R^{-}}=0.
\end{equation}
This last expression (\ref{eq65}), imposes a natural constraint on the free parameter $\alpha$ given by
\begin{equation}\label{eq66}
\alpha<1,    
\end{equation}
in order to preserve $p^{(\text{tot})}_{t}>p^{(\text{tot})}_{r}$ at all points inside the collapsed structure, which in addition ensures  $\Delta>0$, what prevents the system from unwanted behavior such as instabilities. Moreover, from (\ref{eq22}) we obtain
\begin{eqnarray}\label{eq67}
p^{(\text{tot})}_{t}(r)=p^{(\text{tot})}_{r}(r)+\frac{\alpha r^{2}}{C^{2}\left(A^{2}+2r^{2}\right)^{2}\left(A^{2}+3r^{2}\right)^{2}}\bigg[8A^{6}\lambda\nonumber\\
+4A^{4}C^{2}\lambda+24A^{4}r^{2}\lambda+12A^{2}r^{4}\lambda
-24C^{2}r^{4}\lambda\nonumber\\
+12A^{2}C^{2}r^{2}+12C^{2}r^{2}+3A^{4}C^{2}\bigg],
\end{eqnarray}
remembering that $p^{(\text{tot})}_{r}=(1-\alpha)p$.
As it is observed $p^{(\text{tot})}_{t}$ imposes a lower bound on $\alpha$ \i.e, $\alpha>0$. Thus, the positiveness of the total tangential pressure throughout the compact object is ensured. Therefore we have
\begin{equation}\label{eq68}
0<\alpha<1,    
\end{equation}
in order to get a well behaved stellar interior. Another interesting point to be noted here, is that the condition (\ref{eq65}) leads to
\begin{eqnarray}\label{eq69}
C=\frac{1}{2R^{2}+A^{2}-4R^{2}\lambda}\bigg[\bigg(4R^{2}\lambda-2R^{2}-A^{2}\bigg)\bigg(6A^{4}\lambda\nonumber\\
+22A^{2}R^{2}\lambda+24R^{2}\lambda-A^{4}-5A^{2}R^{2}-6R^{4}\bigg)\bigg]^{1/2},
\end{eqnarray}
which shows that $C$ is $\alpha$ independent. Moreover, a detailed computation from (\ref{eq64}) shows that the remaining parameters namely $A$ and $B$ also are $\alpha$ independent when one chooses the constraint (\ref{eq60}) (these expressions are to long to be displayed here, for this reason we only give the appropriate comments). So, expressions (\ref{eq64}) and (\ref{eq69}) are the sufficient and necessary conditions to obtain the full set of constant parameters $A$, $B$ and $C$ describing the interior solution. On the other hand, the remaining thermodynamic observable $\rho^{(\text{tot})}$ can be obtained as follows
\begin{equation}\label{eq70}
\rho^{(\text{tot})}(r)=\rho(r)+\alpha\theta^{t}_{t}(r),   
\end{equation}
where $\rho(r)$ is {given} by Eq. (\ref{eq30}) and $\theta^{t}_{t}(r)$ has the following expression
\begin{equation}
\begin{split}
\theta^{t}_{t}(r)=\frac{1}{C^{2}\left(A^{2}+2r^{2}\right)^{3}\left(A^{2}+3r^{2}\right)^{2}}\bigg[18A ^{10}a +146A^{8}ar^{2}&\\
-20A^{6}C^{2}ar^{2}+522A^{6}ar^{4}-72A^{4}C^{2}ar^{4}+1014A^{4}ar^{6}
&\\-60A^{2}C^{2}ar^{6}+1020A^{2}ar^{8}-24C^{2}ar^{8}+432ar^{10}-3A^{10}&\\
+3A^{8}C^{2}-31A^{8}r^{2}+16A^{6}C^{2}r^{2}-125A^{6}r^{4}+29A^{4}C^{2}r^{4}&\\
-249A^{4}r^{6}
+24A^{2}C^{2}r^{6}-252A^{2}r^{8}+12C^{2}r^{8}-108r^{10}\bigg].
\end{split}
\end{equation}
The anisotropy factor $\Delta$ is given by the following expression 
\begin{eqnarray}\label{eq71}
\Delta(r)=\frac{\alpha r^{2}}{C^{2}\left(A^{2}+2r^{2}\right)^{2}\left(A^{2}+3r^{2}\right)^{2}}\bigg[8A^{6}\lambda +4A^{4}C^{2}\lambda \nonumber\\+24A^{4}r^{2}\lambda+12A^{2}r^{4}\lambda
-24C^{2}r^{4}\lambda \nonumber\\
+12A^{2}C^{2}r^{2}+12C^{2}r^{2}+3A^{4}C^{2}\bigg].
\end{eqnarray}
It should be noted that at the center of the compact configuration $\Delta(0)=0$. This is so because at the center of the star $p^{(\text{tot})}_{t}(0)=p^{(\text{tot})}_{r}(0)$. Besides, $p^{(\text{tot})}_{t}>p^{(\text{tot})}_{r}$ and $p^{(\text{tot})}_{t}>0$ everywhere inside the configuration implying $\Delta>0$ at all points in the inner solution. Fig. \ref{fig1} illustrates the behaviour of the main salient physical quantities that characterize the system. It is worth mentioning that the anisotropic behavior will occur if both pressures are decreasing in nature in the inner region. In fact, the incompatibility compared to the isotropic pressure causes the energy--density to be altered. Obviously, the equilibrium of the system under the action of the gravitational gravity and hydrostatic repulsion is modified. To close this section we would like to mention that this anisotropic solution is not unique. As was pointed out before, different elections and relations on the $\theta$ components and the decoupler function $f(r)$ can be assumed. In the next section, a different {constrain} is considered yielding to a different anisotropic solution. 

\subsection{$\theta$-effects: Mimicking the density for anisotropy}\label{B}

Another way to close the system (\ref{eq34})--(\ref{eq36}) and obtain a physically and mathematically admissible solution, is to consider that the isotropic density given by (\ref{eq30}) mimics its "simile" of the anisotropic sector given by (\ref{eq34}). Then we have
\begin{equation}\label{eq73}
\theta^{t}_{t}(r) =\rho(r).  
\end{equation}
So, by equating Eqs. (\ref{eq30}) and (\ref{eq34}) we arrive to a general expression for the decoupler function $f(r)$ given by
\begin{equation}\label{eq74}
f(r)=\mu(r)-1+\frac{1}{r}\int F_{\lambda}(r)r^{2}dr + \frac{D}{r},   
\end{equation}
being $D$ an integration constant. To avoid divergent behavior in the stellar interior we set $D=0$. Thus (\ref{eq74}) becomes 
\begin{equation}\label{eq75}
f(r)=\mu(r)-1+\frac{1}{r}\int F_{\lambda}(r)r^{2}dr.    
\end{equation}
Thus the deformed radial metric potential $e^{-\eta}$ is given by
\begin{equation}\label{eq76}
e^{-\eta}=\left(1+\alpha\right)\mu(r)+\alpha\left(\frac{1}{r}\int F_{\lambda}(r)r^{2}dr-1\right).    
\end{equation}
Therefore the general deformed Tolman IV solution is written as
\begin{equation}\label{eq77}
\begin{split}
ds^{2}=B^{2}\left(1+\frac{r^{2}}{A^{2}}\right)dt^{2}-\bigg[\left(1+\alpha\right)\mu(r)&\\
+\alpha\left(\frac{1}{r}\int F_{\lambda}(r)r^{2}dr-1\right) \bigg]dr^{2}-r^{2}d\Omega^{2}.   
\end{split}
\end{equation}
As before, the parameters $A$, $B$ and $C$ are obtained from the junction conditions. However, the imposition of constraint (\ref{eq73}) slightly changes the information obtained from condition (\ref{eq46}). Now from (\ref{eq46}) one gets an expression for $C$ in terms of  $\alpha$ parameter. The consequences of this $\alpha$ dependency will be a matter of the following section.

\begin{figure*}
\centering
\includegraphics[width=0.42\textwidth]{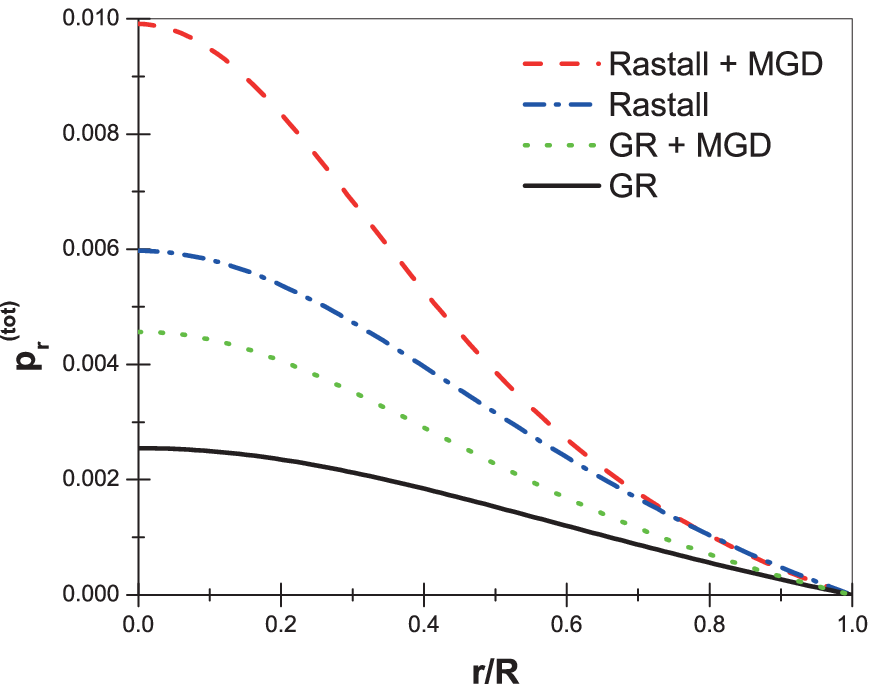}      \~~~
\includegraphics[width=0.42\textwidth]{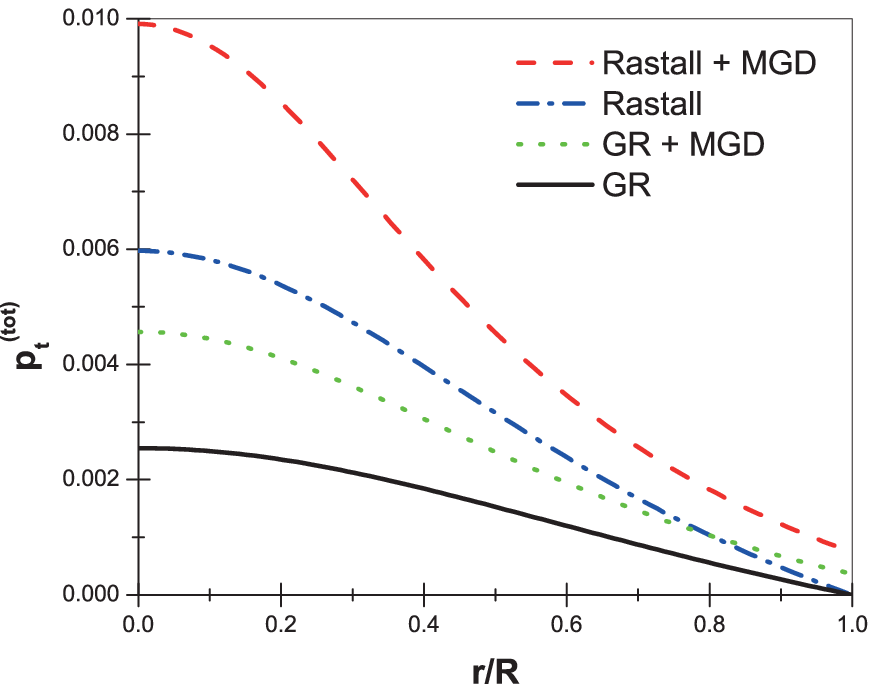}      \
\includegraphics[width=0.42\textwidth]{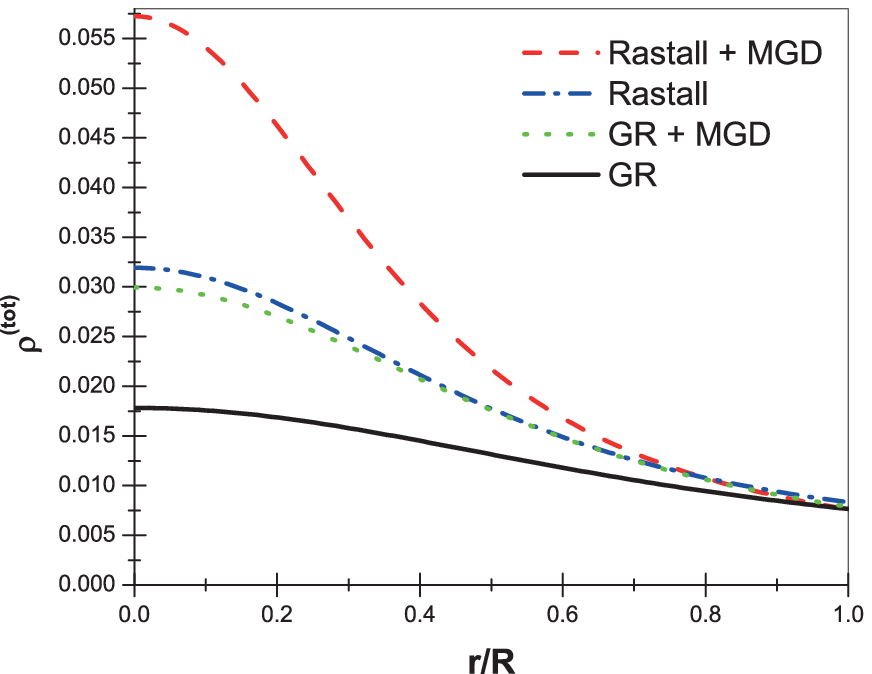}      \~~~
\includegraphics[width=0.42\textwidth]{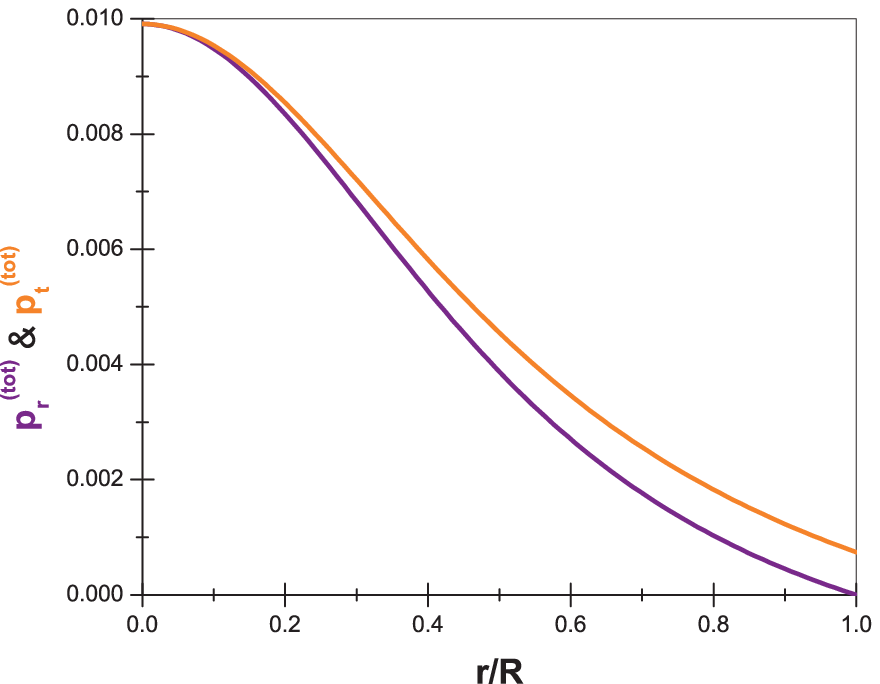}      \
\caption{\textbf{Mimic Constraint} $\rho(r)=\theta^{t}_{t}$. To obtain the trend of the principal thermodynamic observables we have considered throughout the study the following mass-radius ratio $M_{0}/R=0.2$. Moreover, the red curve (dashed) representing Rastall + MGD corresponds to $\alpha=0.2$ and $\lambda=-0.09$, for the blue (dashed--dotted) one corresponding to pure Rastall gravity $\alpha$ and $\lambda$ are $0.0$ and $-0.09$ respectively. Next, the green (short--dashed) line corresponding to GR + MGD takes $\alpha=0.2$ and $\lambda=0.0$. Finally, the black curve (solid) representing GR theory $\alpha=\lambda=0.0$. Upper row: 
{\bf Left panel} illustrates the monotonic behaviour from the center to the boundary of the total radial pressure at all points inside the structure. As it is observed this quantity vanishes at surface. The  {\bf Right panel}: shows the trend of the total tangential pressure everywhere inside the compact object. Lower row: {\bf Left panel} exhibits the behaviour of the total energy--density. Finally, the {\bf Right panel} displays a comparison between the total radial and total tangential pressure. It should be noted that the presence of anisotropies {causes} the pressures values to drift apart.}
\label{fig2}
\end{figure*}

The rest of the principal variables are obtained after inserting the following decoupler function $f(r)$
\begin{equation}\label{eq79}
\begin{split}
f(r)=\frac{1}{8rC^{2}\left(A^{2}+2r^{2}\right)}\bigg[3\sqrt{2}\lambda A\bigg(A^{4}+2A^{2}C^{2}+2A^{2}r^{2}&\\
+4C^{2}r^{2}\bigg)\arctan\left(\frac{r\sqrt{2}}{A}\right)-6A^{4}r\lambda
-12A^{2}C^{2}r\lambda&\\
+8A^{2}r^{3}\lambda
-16C^{2}r^{3}\lambda+32r^{5}\lambda-8A^{2}r^{3}-8C^{2}r^{3}-8r^{5}\bigg]
\end{split}
\end{equation}
into equations (\ref{eq35})--(\ref{eq36}). As it is observed from Eq. (\ref{eq79}), there is a global factor $1/r$. This factor arises in the final expression of the decoupler function $f(r)$ after solve Eq. (\ref{eq73}). In principle, this factor {introduces} a singular behavior at $r=0$. However, this is not a true singularity in the present case, because the $\mathrm{arctan}(r\sqrt{2}/A)$ function is smooth and continuous for all $r$, hence one can {expand} this function in a Taylor series around $r=0$ up to first order in $r$ to eliminated the singular behaviour and keep the dimensionality of all terms. So, the concrete form once the expansion is performed, is given by
\begin{equation}\label{expanded}
f(r)=\frac{\left(5A^{2}\lambda+2C^{2}\lambda+8\lambda r^{2}-2A^{2}-2C^{2}-2r^{2}\right)r^{2}}{2C^{2}\left(A^{2}+2r^{2}\right)}.    
\end{equation}
After that, the thermodynamic observables can be computed as follows 
\begin{eqnarray}\label{eq80}
p_{r}^{(\text{tot})}&=&p(r)-\alpha\theta^{r}_{r} \\ \label{eq81}
p_{t}^{(\text{tot})}&=&p(r)-\alpha\theta^{\varphi}_{\varphi},
\end{eqnarray}
and by virtue of (\ref{eq73}) 
\begin{equation}\label{eq82}
\rho^{(\text{tot})}=\left(1+\alpha\right)\rho(r),  
\end{equation}
where $\rho(r)$ and $p(r)$ are given by Eqs. (\ref{eq30}) and (\ref{eq31}), respectively. The final expressions are too long to be displayed here, for this reason we have omitted them. However, the expression corresponding to the anisotropy factor $\Delta$ is quite small and has the following form
\begin{equation}\label{aniro}
\Delta(r)=\alpha\left(\theta^{r}_{r}-\theta^{\varphi}_{\varphi}\right)=\frac{\alpha\left(3A^{2}\lambda-2C^{2}\lambda+2C^{2}\right)r^{2}}{2C^{2}\left(A^{2}+r^{2}\right)^{2}}.    
\end{equation}
As it {is} observed, the above expression has the usual behaviour \i.e, null at $r=0$ and positive defined everywhere inside the compact configuration iff $\alpha>0$ and if the numerator is also positive defined.
At this stage we have completed our mathematical and graphical analysis with Fig. \ref{fig2}. In the next sections, we are going to discuss in details the affects introduced by MGD on the mass function and compactness factor and also the mathematical and physical implications in the Rastall gravity scenario.

\section{The mass function and  compactness factor}\label{sec7}

In this section we analyze the incidences induced by gravitational decoupling via MGD on the mass function $m(r)$ and compactness factor $u\equiv \frac{m(R)}{R}$. In order to provide a pedagogic explanation, we will start by discussing the implications of MGD in the context of Einstein's gravity theory and then we will move to Rastall's approach. 

As was pointed out in Sec. \ref{sec4}, the MGD process means to deform one of the metric potentials, namely $\nu$ or $e^{-\eta}$. This mechanism allows to separate the seed space--time with {its} matter distribution from the new sector $\theta_{\mu\nu}$, geometrically described by the decoupler function $f(r)$. However, it should be noted that the only way to split the system of equations (\ref{eq27})--(\ref{eq29}) is through the map expressed by Eq. (\ref{eq26}). This is so because, the $t-t$ component of the Einstein field equations only depends on the $g_{rr}$ metric potential and {its} derivative. So, if the MGD is realized on the temporal $g_{tt}$ component of the metric tensor characterized by $\nu$, there is not way to separate $\rho$ from $\theta^{t}_{t}$. Hence, the anisotropic behaviour enters to the system via the radial component of the metric tensor. This entails an important consequence on some of the macro physical observables defining the compact structure, specifically its mass and the associated mass--radius ratio. As it {is} well--known the gravitational mass function can be obtained by direct integration of the $t-t$ component of the Einstein equations, yielding to
\begin{equation}\label{sec71}
m_{GR}(r)=\frac{r}{2}\left[1-e^{-\lambda(r)}\right],  
\end{equation}
which under MGD becomes 
\begin{equation}\label{sec72}
m(r)=m_{GR}(r)+m_{MGD}(r)=\frac{r}{2}\left[1-\mu(r)-\alpha f(r)\right],  
\end{equation}
where we have defined the MGD mass function as follows
\begin{equation}\label{sec73}
m_{MGD}(r)\equiv -\alpha\frac{rf(r)}{2}.
\end{equation}
As it is observed, the original GR mass function $m_{GR}(r)$ is altered by a quantity (\ref{sec73}). In principle, this extra piece induced by MGD grasp could {increase} or {reduce} the mass of the compact structure. Obviously this depends on the sign of $\alpha$ and $f(r)$. Given that these process can occur by different situations, it is very important to keep in mind that there is a mandatory condition to satisfy: a strictly positive and increasing mass function everywhere inside the structure. So, in order to clarify this situation we summarize all the possibilities as follows

\begin{itemize}
    \item \textbf{Case 1:} Positive $\alpha$ and  positive and increasing decoupler function $f(r)$ . In this case the mass of the object is reduced. Nevertheless, this extra piece must grow less than the original mass function $m_{GR}(r)$. 
    \item \textbf{Case 2:} Positive $\alpha$ and negative and decreasing $f(r)$ function. In this opportunity the mass will increase.
    \item \textbf{Case 3:} Negative $\alpha$ and positive and increasing  decoupler function $f(r)$. This case is equal to the case 2. 
    \item \textbf{Case 4:} Negative $\alpha$ and negative and decreasing deformation function $f(r)$. This case is equal to the case 1. 
\end{itemize}

\begin{figure}
\centering
\includegraphics[width=0.237\textwidth]{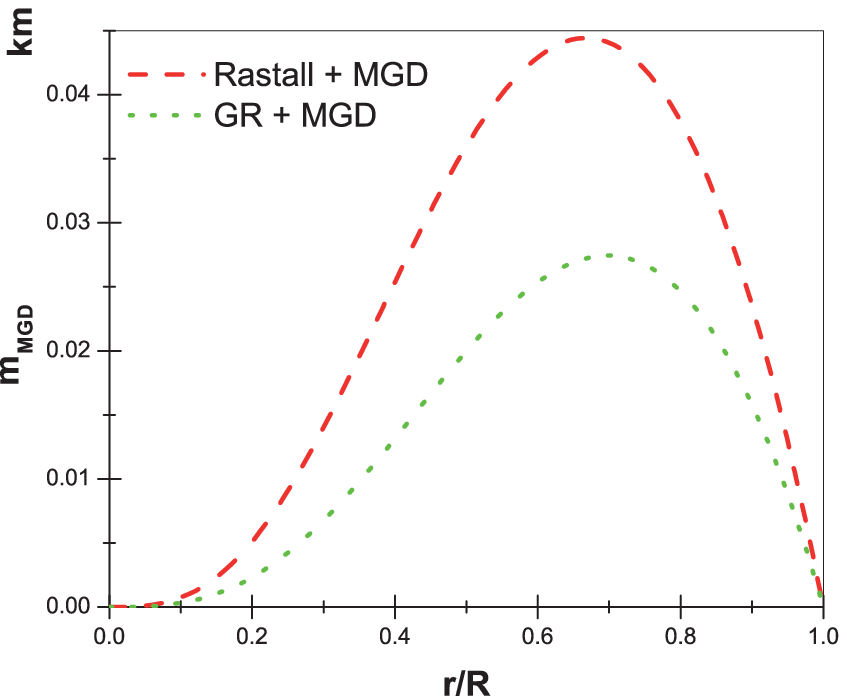}    
\includegraphics[width=0.237\textwidth]{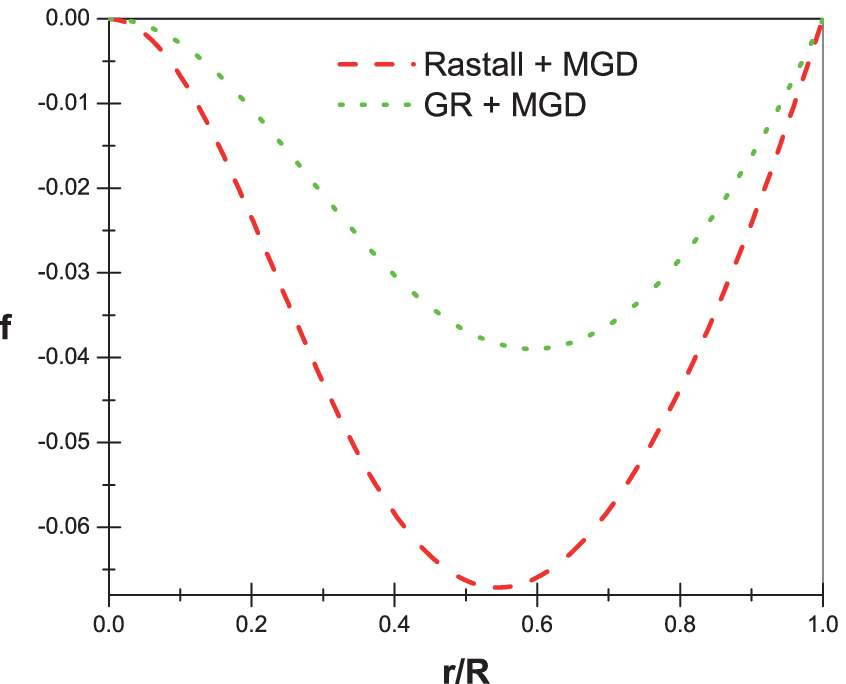}     
\includegraphics[width=0.237\textwidth]{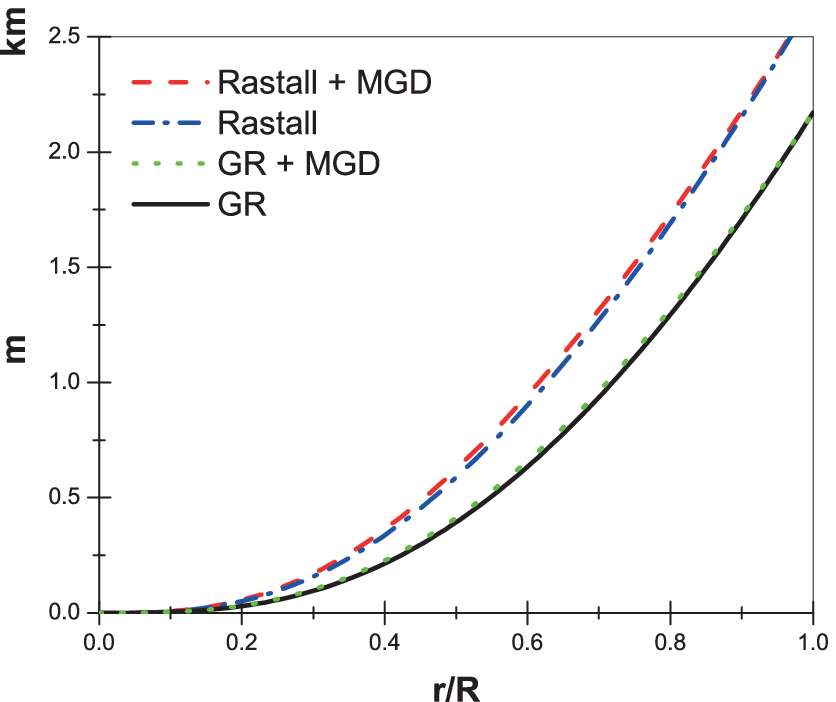} 
\includegraphics[width=0.237\textwidth]{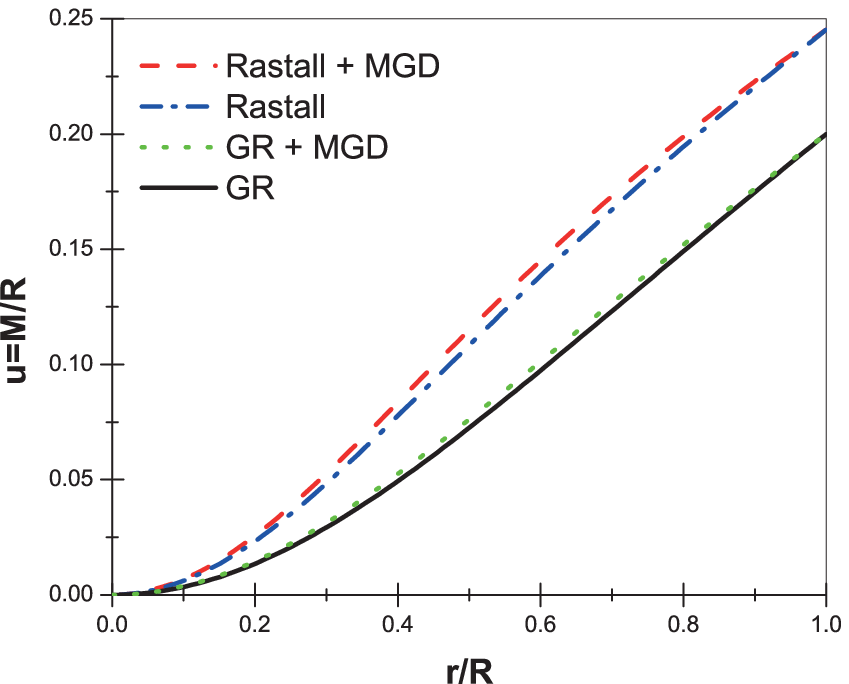}      
\caption{\textbf{Mimic Constraint} $p(r)=\theta^{r}_{r}$. The \textbf{top left} panel shows the MGD mass function (\ref{sec73}) versus radial coordinate $r/R$ while the \textbf{right panel} represents the deformation function $f(r)$ against the radial coordinate $r/R$ corresponding to Rastall+MGD (dashed curve) and GR+MGD (dotted curve). We observe from figures (top one) that the value of the MGD mass function is same at the boundary $r=R$ in both Rastall and GR theories, because the deformation function vanishes at the boundary $r=R$.  
{\bf bottom left} figure illustrates mass function (\ref{sec711}) inside the stellar structure for the Rastall+MGD (dashed), Rastall (dot--dashed), GR+MGD (dotted), and GR (solid) scenario. The \textbf{bottom right} figure shows the compactness $u\equiv\frac{m(r)}{r}$ versus radial coordinate $r/R$ with the same description the curve as in left panel. As we see, clearly the mass function $m(r)$ and compactness $u$ coincide at the boundary $r=R$ for Rastall+MGD and pure Rastall case. This situation also occurs for GR+MGD and GR scenarios. This happens due to the no contribution of MGD mass at the boundary i.e. $m_{MGD}(R)=0$. }
\label{fig3}
\end{figure}

It is obvious that the above general analysis, is valid for all $r$ belonging to the interval $[0,R]$. Now, the sign of the constant $\alpha$ depends on several factors. The most important is associated with the anisotropy factor, which in general if the seed solution is described by a perfect fluid matter distribution is defined by (\ref{eq22}). So, if $\theta^{r}_{r}>\theta^{\varphi}_{\varphi}$ then $\alpha$ must be positive, otherwise $\alpha$ is negative. On the other hand, the behaviour of the decoupler function $f(r)$ is subject to the closure of the $\theta$--sector. As was pointed out earlier, to close the problem at least from the mathematical point of view it is necessary to supplement with additional information. For example in \cite{r69,r70,r91}, the  decoupler function $f(r)$ was specified without assuming any relation between the seed and $\theta_{\mu\nu}$ {sector} or by imposing some {constraints} on the $\theta_{\mu\nu}$ components. Thus, the behaviour of $f(r)$ in the mentioned {cases}, is only determined by the behaviour of the $\mu(r)$ metric potential \i.e, positive defined and strictly increasing function with increasing $r$ within the star. In the present study the situation is quite different because the deformation function $f(r)$ is obtained by imposing the so--called mimic constraint, Eqs. (\ref{eq60}) and (\ref{eq73}). In this respect, the first of these restrictions involves an interesting situation. When the $\theta^{r}_{r}$ component is mimicking the isotropic pressure ($p(r)$), then the decoupler function $f(r)$ has the general form
\begin{equation}\label{sec74}
p(r)=\theta^{r}_{r}(r)\Rightarrow f(r)=\frac{1}{1+r\nu^{\prime}(r)}-\mu(r).   
\end{equation}
From the $r-r$ component of the Einstein field equations 
\begin{equation}\label{sec75}
p(r)=-\frac{1}{r^{2}}+\mu(r)\left(\frac{\nu^{\prime}(r)}{r}+\frac{1}{r^{2}}\right),    
\end{equation}
as the pressure in the radial direction must be vanish at the boundary $\Sigma\equiv r=R$, from (\ref{sec75}) one gets
\begin{equation}\label{sec76}
R\nu^{\prime}(R)=\frac{1}{\mu(R)}-1.   
\end{equation}
Next, evaluating $f(r)$ at $r=R$ from expression (\ref{sec74}) and replacing (\ref{sec76}) one arrives to 
\begin{equation}\label{sec77}
f(R)=0. 
\end{equation}
This implies that the total mass $m_{GR}(R)+m_{MGD}(R)$ contained by the sphere and observed by a distant observer coincides with the original mass $m_{GR}(R)$. This is a general result independent of the theory. Now, in considering the present case, the Rastall mass is also not modified by MGD. To see this, we obtain the total mass function $m(r)$ from Eqs. (\ref{eq41})--(\ref{mlambda}) as 
\begin{equation}\label{sec78}
\begin{split}
m(r)=\frac{r}{2}\left[1-\mu(r)\right]\left[1-4\lambda\right] -\alpha\frac{r}{2}f(r)
&\\  +\frac{3\lambda}{2}\int r\left[\mu(r)\nu^{\prime}(r)-\mu^{\prime}(r)\right]dr.  
\end{split}
\end{equation}
It is evident that when $\alpha=\lambda=0$ the GR mass function is recovered, what is more if $\lambda=1/4$ the mass function in not well defined. This reinforce the previous comments about this forbidden value for the Rastall parameter. Now, as before, evaluating Eq. (\ref{eq31}) at the surface one obtains
\begin{equation}\label{sec79}
R\nu^{\prime}(R)=\frac{1-4\lambda+\lambda r\mu^{\prime}(R)+4\lambda \mu(R)-\mu(R)}{\mu(R)\left(1-3\lambda\right)},   
\end{equation}
subject to $\lambda\neq 1/3$. Then, replacing the above result in Eq. (\ref{eq61}) evaluated at the boundary, after some algebra one gets again
\begin{equation}\label{sec710}
f(R)=0.    
\end{equation}
So, in this case ($p_r=\theta^r_r$) the mass of the compact configuration is given by the GR gravitational mass plus Rastall contribution. The mass function $m(r)$ associated to the model under study is given by
\begin{eqnarray}\label{sec711}
m(r)=\frac{r}{2}\left[1-\frac{\left(1-\frac{r^{2}}{C^{2}}\right)\left(1+\frac{r^{2}}{A^{2}}\right)}{\left(1+2\frac{r^{2}}{A^{2}}\right)}\right]\bigg[1-4\lambda\bigg]\nonumber \\
+\frac{\alpha r^{3}}{2C^{2}\left(A^{2}+2r^{2}\right)^{2}\left(A^{2}+3r^{2}\right)}\bigg[\bigg(A^{2}+r^{2}\bigg) \nonumber\\
\times\bigg(6A^{4}\lambda
+22A^{2}r^{2}\lambda-4C^{2}r^{2}\lambda+24r^{4}\lambda
\nonumber\\-5A^{2}r^{2}+2C^{2}r^{2}-6r^{4}-A^{4}+A^{2}C^{2}\bigg)\bigg]  \nonumber\\
+\lambda\frac{3 \left(A^{2}+2C^{2}\right)r^{3}}{4C^{2}\left(A^{2}+2r^{2}\right)}.~~~
\end{eqnarray}
Therefore, if the mass remains unchanged when the isotropic pressure is mimicked the same occurs with the compactness factor $u$, since it depends on the total mass $M$. To support the previous discussion Fig. \ref{fig3} shows the behaviour of the MGD mass and the decoupler function inside the star (upper panels). As can be seen both $m_{MGD}(r)$ and $f(r)$ are vanishing at the boundary of the configuration implying that there is not contribution coming for the MGD sector to the total mass of the object. In fact. this is the peculiarity of the constraint (\ref{eq60}), the mass inside the star is redistributed around the core while towards the boundary is attenuated. One can confirm this by contrasting the Rastall+MGD and pure Rastall density profiles given in Fig. \ref{fig1} (left lower panel), where it is observed that the density is greater in the Rastall+MGD scenario than in the pure Rastall case at some point before  reaching the surface, where the Rastall+MGD curve (red) has a discontinuity \i.e, is dominated by the pure Rastall scenario. The other test that shows that the mass is only redistributed within the structure is reflected in Fig. \ref{fig3} (lower panels) which clearly illustrates that both the mass function $m(r)$ and the compactness parameter as a function of the radial coordinate $u(r)$ remain unchanged in both domains: Rastall+MGD and pure Rastall. Besides, we have displayed in the same plots the GR and GR+MGD cases to prove that this is a general result, independent of the underlying theory. To finalize the discussion regarding the mimic constraint (\ref{eq60}), it should be noted that the mass function curve in the Rastall+MGD and pure Rastall setting coincide at $r=0$ and $r=R$ while from $r/R=0.3$ to $r/R=0.9$ (approximately) are slightly jarring (the same occurs with the compactness parameter), this difference also shows the mass displacement within the object around the core. \\
%%%%%%%%%%%%%%%%%%%%%%%%%
On the other hand, the situation is quite different when \textbf{$\theta^{t}_{t}$ mimics the seed density $\rho (r)$} (\ref{eq73}). In this case, the mass function $m(r)$ associated with this constraint is given as
\begin{equation}\label{sec712}
\begin{split}
m(r)=\left(1+\alpha\right)\Bigg(\frac{r}{2}\left[1-\frac{\left(1-\frac{r^{2}}{C^{2}}\right)\left(1+\frac{r^{2}}{A^{2}}\right)}{\left(1+2\frac{r^{2}}{A^{2}}\right)}\right]\bigg[1-4\lambda\bigg]&\\
+\lambda\frac{3 \left[A^{2}+2C^{2}\right]r^{3}}{4C^{2}\left[A^{2}+2r^{2}\right]}\Bigg).
\end{split}
\end{equation}

\begin{figure}
\centering
\includegraphics[width=0.237\textwidth]{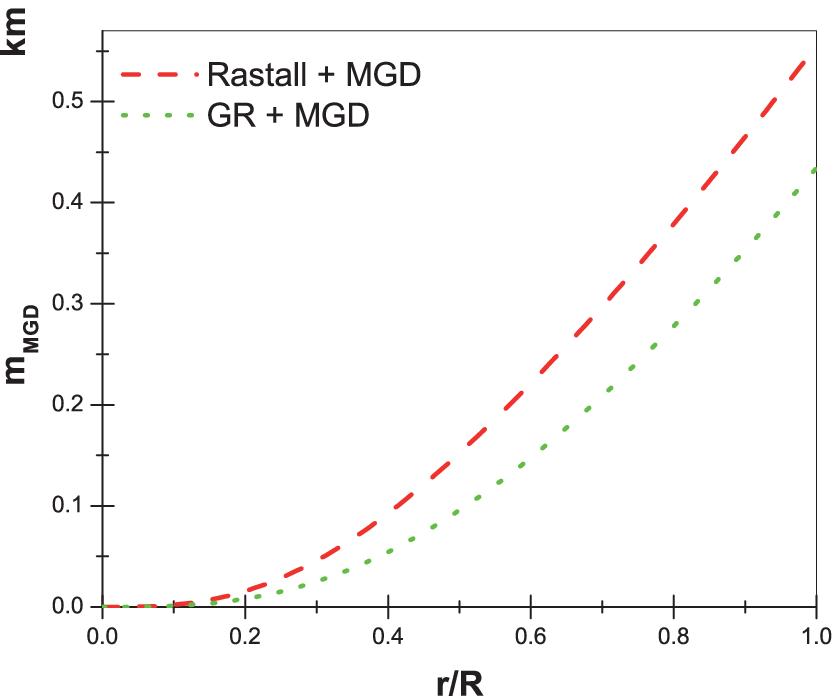}
\includegraphics[width=0.237\textwidth]{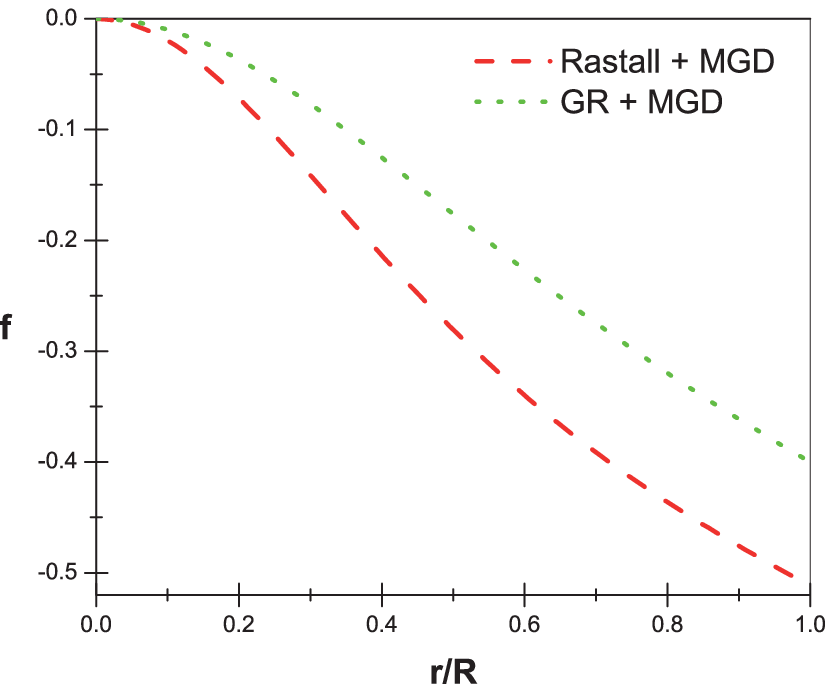}  
\includegraphics[width=0.237\textwidth]{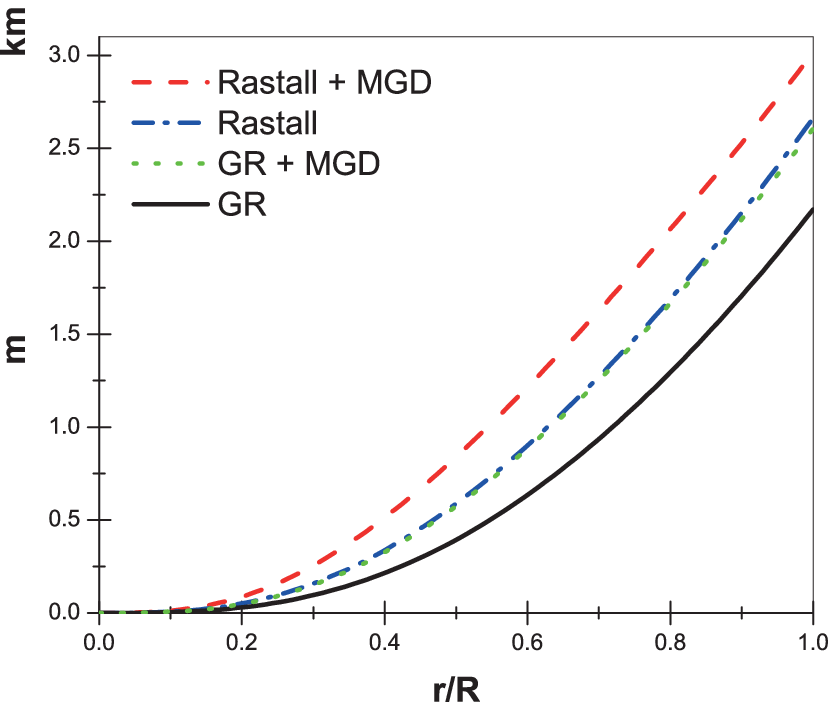}  
\includegraphics[width=0.237\textwidth]{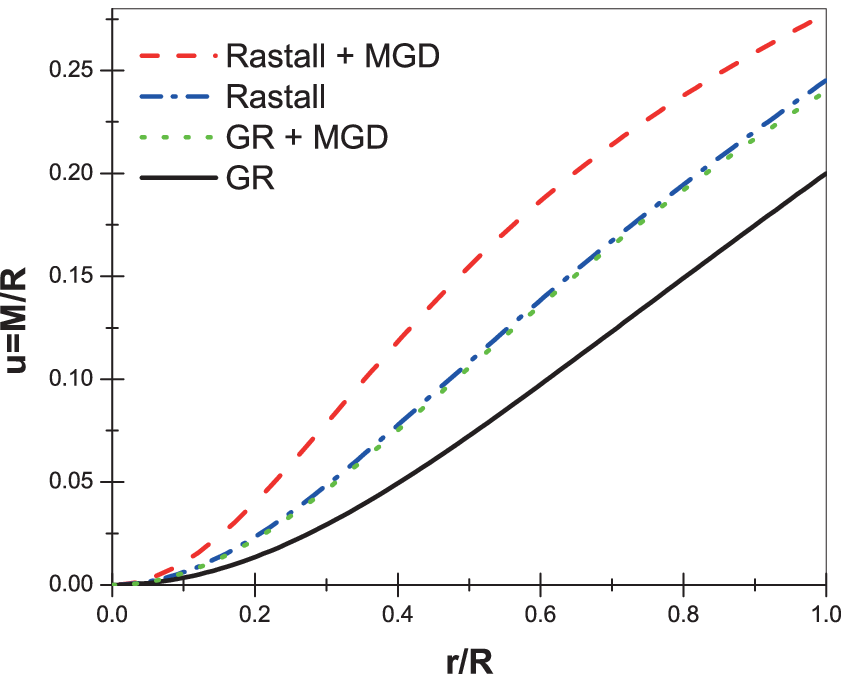}      
\caption{\textbf{Mimic Constraint} $\rho(r)=\theta^{t}_{t}$. The \textbf{top left} panel shows the MGD mass function (\ref{sec73}) versus the radial coordinate $r/R$, while the \textbf{right panel} represents the deformation function $f(r)$ versus $r/R$ corresponding to Rastall+MGD (dashed curve) and GR+MGD (dotted curve). Here the situation is different than the previous mimic constraint. The MGD mass is increasing monotonically throughout within the stellar object and it has greater value in Rastall case, which is happening due to the trend of deformation function $f(r)$ within the object.  
{\bf bottom left} figure illustrates mass function (\ref{sec712}) inside the stellar structure for the Rastall+MGD (dashed), Rastall (dot-dashed), GR+MGD (dotted), and GR (solid) {scenarios}. The \textbf{bottom right} figure shows the compactness $u\equiv\frac{m(r)}{r}$ against the dimensionless radial coordinate $r/R$ with the same description of the curve as in left panel. It is clear from both bottom figures that the mass $m(r)$ and compactness $u$ in Rastall+MGD case is larger than the pure Rastall, GR+MGD and pure GR {cases} within the stellar structure. However, $m(r)$ and $u(r)$ have approximately the same value in pure Rastall and GR+MGD within the object.} 
\label{fig4}
\end{figure}

As Fig. \ref{fig4} demonstrates, the usage of this restriction leads to an increase in the mass of the compact object (see left panel in the lower row). This is so because the total mass $M=m(R)$ is proportional to the total seed mass $M_{0}+M_{\lambda}$ by a factor $(1+\alpha)$. Of course, {this} happens since $\alpha>0$ and $f(r)<0$ (Fig. \ref{fig4} right panel in the upper row ), otherwise the mass will decrease. This case {corresponds} to mentioned case 1 above. As the mass increases in both scenarios Rastall + MGD and GR + MGD, then the compactness factor also increases (right lower panel in Fig. \ref{fig4}). Remembering that we have fixed $u=M_{0}/R=0.2$, it is clear that the MGD contribution always arises as a mass generator, taking into account that this strongly depends on the sign of $\alpha$ and $f(r)$.   

%%%%%%%%%%%%%%%%%%%%%%%%%
\subsection{Bounds induced on $\alpha$ and $\lambda$ by observational data }

In considering that the previous results depend on the choice made on $\alpha$ and $\lambda$, in order to provide a more real picture and reliable model we bound the mentioned parameters by using real observational data. {In this opportunity we have selected the well--known millisecond pulsar PSR J1614--2230. The mass of this neutron star is $1.97\pm0.04 M_{\odot}$ and corresponds to the highest precisely
measured neutron star mass determined to date. It is worth mentioning that to determine the radius $R$ of these structures, it strongly depends upon two main ingredients: i) the equation of state (EoS) driven the interaction inside the star, and ii) the method employed to observe the object, for example X--rays, Shapiro delay, etc. In this regard, the EoS could describe the interaction of nucleons, nucleons coupled to exotic hadronic matter
such as hyperons or kaon condensates and strange matter (formed by u, d and s quarks), to name a few. On the other hand, the method could or not provide information about the size of the configuration \i.e, its radius $R$. For example, the measures based on X--rays provide information on the mass--radius ratio of the object, thus the radius $R$ can be predicted. However, the Shapiro delay provides no information
about the neutron star radius. In this case to fix the radius $R$ and explore the limit of the theory, we shall follow the information given in \cite{demorest}. In that work, the radius $R$ of the  millisecond pulsar PSR J1614--2230, can be inferred by looking at the M--R curve. In this concern, Demorest et al.\cite{demorest} used the Shapiro delay approach together with different EoS, thus determining which interaction leads to obtaining the reported mass value for PSR J1614--2230. What is more, when nucleon interaction is considered the radius range is $11-15\ [\text{km}]$, while for the strange matter, it is $10-11\ [\text{km}]$ (for further details see Fig. 3 in \cite{demorest}).      }  

{Before starting to discuss about the boundedness of the parameters $\alpha$ and $\lambda$, we would like to highlight some it is important to mention that we shall concentrate the analysis only in the mimic constraint given by Eq. (\ref{eq73}). As was discussed previously, the restriction (\ref{eq60}) only causes a mass redistribution inside the compact structure. Then the total mass $M$ and its associated compactness parameter $u$ are not altered. Besides, $\alpha$ is automatically bounded by above and below as shown Eq. (\ref{eq68}). To understand how these parameters are disturbed under the mimic constraint (\ref{eq73}). For this purpose, we start revisiting the GR+MGD scenario and then the pure Rastall case. Finally, the Rastall+MGD framework will be discussed.}

{Since any affection on the total mass $M$ is reflected in the compactness factor $u$, it is better to deal with $u$ instead of $M$. This is so because $u$ tell us how the compact the object is. So, by virtue of Eq. (\ref{eq82}) one gets
\begin{equation}\label{news1}
M=\left(1+\alpha\right)M_{0},  
\end{equation}
remembering that the subscript $0$ stands for pure GR scenario. Then, 
\begin{equation}\label{news2}
u=\left(1+\alpha\right)u_{0},    
\end{equation}
where the condition $u/u_{0}>1$ implies $\alpha>0$.
On the other hand, for isotropic fluid spheres the Buchdahl limit in the context of GR says that $u_{0}\leq4/9$. The extreme case $u_{0}=4/9$ in conjunction with $\alpha>0$ establishes a lower bound, being the upper one $u_{BH}=1/2$, that is, the black hole limit. Of course, a compact object whose mass--radius ratio corresponds to $4/9$ has not yet been reported. What is clear, is the fact that MGD allows for more compact structures within the context of GR. Following the same spirit, in the framework of pure Rastall gravity, the mass and compactness factor are also altered, by the non--trivial contribution introduced by the Rastall sector. In this case the mass--radius ratio reads
\begin{equation}\label{news3}
u=u_{0}\left(1-4\lambda\right)+\frac{3\lambda}{2R}\int^{R}_{0}r\left(\mu\nu^{\prime}-\mu^{\prime}\right)dr.    
\end{equation}
Then, to assure $u>u_{0}$ one requires
\begin{equation}\label{news4}
\lambda\left[\frac{3}{2R}\int^{R}_{0}r\left(\mu\nu^{\prime}-\mu^{\prime}\right)dr-8u_{0}\right]>0.    
\end{equation}
It is evident that to satisfy the above restriction, the sign of $\lambda$ depends on the sign of the bracket, what is more the result of the  integral term depends on the choice of the metric potentials. Now if MGD is incorporated and taking into account (\ref{eq82}), equation (\ref{news3}) becomes
\begin{equation}\label{news5}
u=\left[1+\alpha\right]\left[u_{0}\left(1-4\lambda\right)+\frac{3\lambda}{2R}\int^{R}_{0}r\left(\mu\nu^{\prime}-\mu^{\prime}\right)dr\right].     
\end{equation}
As can be seen $\alpha$ and $\lambda$ are quite involved. So in general, it is not an easy task to determine the magnitude and sign of these parameters, in order to have $u>u_{0}$. But for this specific situation, where $\nu$ and $\mu$ are given by (\ref{eq56})--(\ref{eq57}), the total mass $M$ reads
\begin{equation}
\label{news6}
M\simeq\left[1+\alpha\right]\left[\frac{2\left(A^{2}+C^{2}+R^{2}\right)-\lambda\left(7A^{2}+6C^{2}+8R^{2}\right)}{4C^{2}\left(A^{2}+2R^{2}\right)}\right]R^{3},    
\end{equation}
and the expression (\ref{news5}) becomes
\begin{equation}\label{news7}
u\simeq \left[1+\alpha\right]\left[u_{0}\left(1-4\lambda\right) +\lambda\frac{3 \left(A^{2}+2C^{2}\right)R^{2}}{4C^{2}\left(A^{2}+2R^{2}\right)}\right].
\end{equation}}

{As was pointed out before, $\alpha$ and $\lambda$ are too much involved to be bounded separately. So, to explore the limit of the theory \i.e, Rastall+MGD, we shall proceed in  the same way as proposed by Linares et al. \cite{linares}. In that article, the authors studied the impact of Weyl contributions inside and outside of the star in the framework of the brane world. The maximum compactness factor was obtained by analyzing the total mass $M$ behaviour against the radius $R$ and by considering different orders of magnitude for the brane world parameter. So, following the same spirit, we take different orders of magnitude for $\lambda$ to elucidate the limit of the theory under this particular solution. At this point some comments are in order. First, to assure $M>0$ from Eq. (\ref{news6}) one has
\begin{equation}\label{news8}
\lambda<\frac{2\left(A^{2}+C^{2}+R^{2}\right)}{7A^{2}+6C^{2}+8R^{2}}.    
\end{equation}
As can be seen, the maximum order of magnitude of the right member in (\ref{news8}) is $10^{-1}$. This bound is independent of the values that $A$, $C$ and $R$ take. Second, although $\alpha> 0$, it cannot be arbitrarily large, since the extreme limit for the compactness factor corresponds to the black hole one \i.e, $u_{BH}=1/2$. Then from (\ref{news2}), it is evident that the exceeding order of magnitude more than $10^{-1}$ can create an unrealistic situation for some $u_0$. Thus, plausible lower and upper bounds for $\alpha$ are  \begin{equation}\label{news9}
  0<\alpha\leq 10^{-1}.  
\end{equation}
So, as it is appreciated in Fig. \ref{fig5} the maximum total mass $M$ and compactness factor $u$ (purple curve) correspond to the lowest value of $\lambda$ (see table \ref{table1}). Respect to the compactness factor, its value is bounded by the extreme GR isotropic case, that is, $4/9$. However, $u=0.367612$ is beyond the usual data reported in the literature within the framework of pure GR in dealing with isotropic/anisotropic compact objects. Besides, the total mass $M=2.919$ is also above the apprised experimental data. Hence, there is a clear tendency to increase the total mass and mass--radius ratio values when $\lambda$ changes both its magnitude and sign.

\begin{figure}
\centering
\includegraphics[width=\columnwidth]{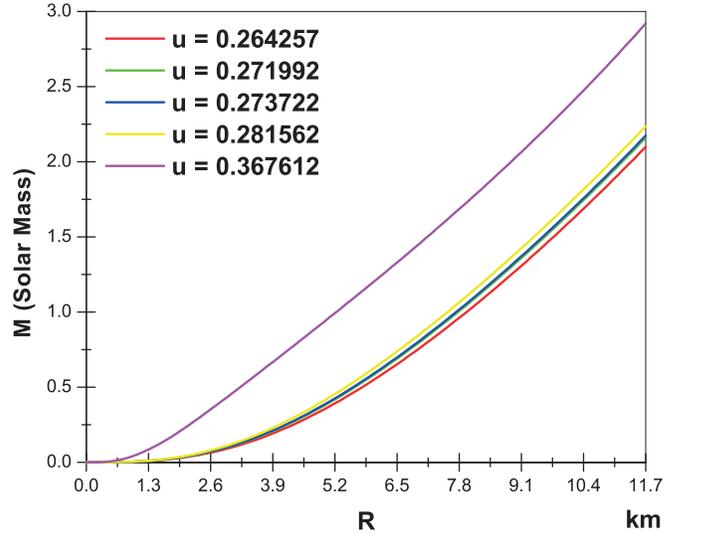}
\caption{The M--R curve for different values of $\lambda$ mentioned in table \ref{table1} and $\alpha=0.1$. See
text for more details.
}
\label{fig5}
\end{figure}
\begin{table}
\caption{The total mass $M$ and compactness factor $u$ for different values of $\lambda$, $R=11.7 \ [\text{km}]$, $\alpha=0.1$, $M_{0}=1.97 M_{\odot}$ and $u_{0}=0.248052$.  }
\label{table1}
\begin{tabular*}{\columnwidth}{@{\extracolsep{\fill}}lrrrrrrrl@{}}
\hline
\multicolumn{1}{c} {$\lambda$}&  
\multicolumn{1}{c} {$M/M_{\odot}$}& 
\multicolumn{1}{c} {$u$}\\
\hline
$10^{-2}$  &$2.099$&$0.264257$  \\
\hline
$10^{-3}$  &$2.160$&$0.271992$  \\
\hline
$-10^{-3}$  &$2.174$&$0.273722$ \\
\hline
$-10^{-2}$&$2.236$&$0.281562$
\\ 
\hline
$-10^{-1}$&$2.919$&$0.367612$\\
\hline
\end{tabular*}
\end{table}

\section{Analysis and Discussion}\label{sec8}
In this section we analyze the physical consequences of the results obtained in \ref{A} and \ref{B}. So, the pertinent comments for the resulting interior solution obtained in \ref{A} are :
\begin{enumerate}
    \item Regarding the junction conditions. The first fundamental form carries an interesting conclusion about the extra piece introduced by the decoupler function in the radial metric potential $e^{-\eta}$. In performing the matching conditions with the vacuum exterior space--time described by Schwarzschild solution (see Eq. (\ref{eq64})), the geometric sector describing the deformed part is totally vanished. This means that the total mass contained by the sphere seen by a distant observer matches the original mass of the object \i.e, $m(R)=M_{0}+M_{\lambda}$. An important point to be noted is that for $r<R$ the mass function ${m}(r)$ carries the MGD contribution. This is because
    the energy--density has and extra contribution coming from the $\theta$--sector. Therefore, the seed energy--density $\rho$ is altered by the presence of this additional contribution. Nonetheless, the mass remains the same with or without $\theta$--sector. From the physical point of view, as it is shown by the left panel of  Fig \ref{fig1} (lower row) in the framework of Rastall+MGD (red curve) the total energy--density dominates other scenarios. The object is denser at the center in the Rastall+MGD background, but towards the surface, the total energy--density is dominated by other scenarios. This behavior suggests that the same mass is redistributed around the core of the star and diminishing towards the surface of the object. 
    \item From the second fundamental form, explicitly given by Eq. (\ref{eq65}) one gets and expression for the constant $C$ independent of the free parameter $\alpha$. this implies that $C$ takes the same values in considering pure Rastall gravity and Rastall+MGD approach. Moreover, as $p_{r}^{(\text{tot})}=\left(1-\alpha\right)p(r)$, the resulting central pressure is below the central pressure in Rastall gravity as illustrates Fig. \ref{fig1} in the upper row (left and right panels). This $\alpha$--independence is a feature of the mimic constraint (\ref{eq60}). In addition as $\alpha$ is restricted to belong to $(0,1)$ in order to obtain an admissible interior solution, by imposing (\ref{eq60}) the resulting object's core will be denser when adding MGD approach to other theories and the central pressure will be less than the seed central pressure.   
    \item Local anisotropies arising in the system due to the presence of the extra source $\theta_{\mu\nu}$, introduce a positive anisotropy factor $\Delta(r)$ at all points inside the compact structure. This is a very important issue in the study of compact configurations because a positive anisotropy factor introduces a repulsive force that counteracts the gravitational gradient. Therefore preventing the compact object from collapsing below its Schwarzschild radius. In addition stability and balance mechanisms are enhanced \cite{r102,r103}. Besides,  as was pointed out by Gokhroo and Mehra \cite{r104} a positive anisotropy factor allows to build more compact objects. In the present study, this feature is depicted in the right panel (lower row) of Fig. \ref{fig1}. It is observed that both the total radial and total tangential pressures coincide at the center and then drift apart towards the boundary of the object. 
    \item Another relevant point is the macro information of the compact structure \i.e, the mass and radius, obtained from astrophysical observations. Both quantities are related by means of the compactness factor $u$, which is related with the surface gravitational red--shift $z_{s}$. Explicitly
    \begin{equation}
    z_{s}=\frac{1}{\sqrt{1-2u}}-1.    
    \end{equation}
    In considering isotropic fluid spheres $u$ has and upper bound known as Buchdahl's limit \cite{r105} given by $u\leq 4/9$. By taking the equality, the maximum allowed gravitational surface red--shift for an isotropic spherical matter distribution is $z_{s}=2$. Nevertheless, Ivanov studies \cite{r106} suggested that when anisotropies are included in the stellar interior, the surface gravitational red--shift increases its maximum value in comparison with its isotropic counterpart. Obviously, $z_{s}$ can not be arbitrarily large and its maximum value depends on the mechanism to introduce anisotropies into the system. In this respect as we discussed above, the mimic constraint (\ref{eq60}) does not alter the total mass (keeping the same radius) of the configuration, only is redistributed within the object. So, in this case, the observed $z_{s}$ does not change despite the system contains local anisotropies \cite{r57,r73}. The reasons behind this behavior is due to the MGD contribution encode in the decoupler function $f(r)$ is vanishing at the boundary $r=R$ of the structure. Therefore the original total mass of the object $M=M_{0}+M_{\lambda}$ remains unaltered. This is corroborated in the upper panels in Fig. \ref{fig3}, as can be seen the MGD mass has not the usual strictly increasing behaviour with increasing radius, it has two identical minimum values ($m_{MGD}=0$) at $r=0$ and at $r=R$. This is supported by the upper right  panel where, it is evident that $f(R)=0$. Besides, the mass function and compactness parameter illustrated in the lower panels of Fig. \ref{fig3} certify that for Rastall and Rastall+MGD, the mass function and the mass--radius ratio are exactly the same (to reinforce this point we have added the GR and GR+MGD cases to show that this result is independent of the theory).
\end{enumerate}

Now we proceed with the appropriated comments for the results obtained in \ref{B}. In this respect, the mimic constraint (\ref{eq73}) 
gives more interesting results than the mimic constraint (\ref{eq60}).

\begin{enumerate}
    \item By imposing the mimic constraint (\ref{eq73}), the resulting junction conditions provide new insights in considering the total mass of the compact object. This time the extra piece $f(r)$ contributes to the matching conditions. This means that the coupling constant $\alpha$ has an active role. So, the observed mass is no longer the same. This is so because 
     \begin{eqnarray}
     m(r)=4\pi\int \rho^{(\text{tot})}r^{2}dr=4\pi\int \left(1+\alpha\right)\rho r^{2}dr, ~~~ 
    \end{eqnarray}
    then
    \begin{equation}
    m(r)=\left(1+\alpha\right)\left(m_{GR}+m_{\lambda}\right)(r).  
    \end{equation}
    Therefore, by virtue of (\ref{eq73}) the mass function $m(r)$ is proportional to the mass of the seed solution given by $m_{GR}+M_{\lambda}$. However, the maximum value that the mass and energy--density of the compact structure can take, is strongly constrained by the values taken by the dimensionless constant $\alpha$. This is because parameter $C$ now depends on $\alpha$ and $\lambda$. This is evident from the condition of null pressure on the surface of the object, which is different from the case previously considered where the constant $C$ only depended on the Rastall parameter $\lambda$. The behavior of the constant $C$, in this case, determines the energy--density behavior at the center of the object and consequently the value of its mass. So, to obtain a denser object, we need to go in the direction of increasing energy--density and mass. This is possible by assigning small values to parameter $\alpha$, which implies that the constant $C$ decreases in module. Nevertheless, the value of parameter $\alpha$ cannot be arbitrarily small since the effects of anisotropy on the stellar interior would be negligible. In addition, negative values of $\alpha$ would introduce instabilities in the system since the total tangential pressure would be less than the total radial pressure, which represents a physically inadmissible situation.
    \item With respect to the central pressure, it increases if the magnitude of $\alpha$ decreases. If $\alpha$ is very small (close to zero) the anisotropy from $\theta$--sector will be negligible. If $\alpha$ is negative then the anisotropy factor $\Delta$ will be too, introducing into the system a force attractive in nature. In conclusion, $\alpha$ is bound to be positive defined. However, it should be noted that the restrictions imposed by choice (\ref{eq73}) on the parameter $\alpha$ depend on the chosen seed solution. For example, for the Heintzman IIa \cite{r71} and Durgapal-Fuloria \cite{r73} isotropic models, studied in the framework of GR+MGD, the mimic constraint (\ref{eq73}) only allows negative values for $\alpha$, which ensures a physically acceptable solution. 
    
    \item  Finally, it is important to highlight that in the present case the observational differences between isotropic and anisotropic distributions are evident. Due to the total mass varies then compactness factor $u$ changes. This fact alters the surface gravitational red--shift $z_{s}$ value. It follows immediately from the definition of $z_{s}$
    \begin{equation}
    z_{s}(\alpha)= \frac{1}{\sqrt{1-2u(\alpha)}}-1,  
    \end{equation}
    where the $\alpha$ dependency is explicit, if $\alpha$ increases then the mass grows, in consequence $u$ increases. Hence, the factor $1/\sqrt{1-2u(\alpha)}$ increases implying that $z_{s}$ grows its value as it is expected when the compact object becomes denser. To verify the impact of MGD on the main macro physical observables by using (\ref{eq73}) we have performed the same analysis as before. As Fig. \ref{fig4} illustrates, the $m_{MGD}$ mass (upper left panel) behaves as usual \i.e, increasing function with increasing radial coordinate, reaching its maximum value at the surface.  On the other hand, the deformation function $f(r)$ (upper right panel) has a decreasing behaviour with in conjunction with a positive $\alpha$  provides in principle a total mass increased by a certain amount (it should be noted that the total mass also depends on $\lambda$). The lower panels sketched the mass function and mass--radius relation, where the MGD effects on these quantities are evident.
\end{enumerate}
To conclude this section 7.1, it is important to highlight that we have explored the limits of the theory under the assumption of this particular model, by bounding the free parameters $\alpha$ and $\lambda$ with the help of real observational data. For the limits we mean the maximum and minimum order of magnitude of $\alpha$ and $\lambda$ yielding to a reasonable results from the astrophysical point of view \i.e, plausible values for the total mass $M$ and mass--radius ratio $u$. The analysis was performed by considering the restriction (\ref{eq73}) only. This consideration is based on  earlier discussions about how mimetic constraints modify the main features of the system. As illustrated by table \ref{table1} and Fig. \ref{fig5}, in the scenario of Rastall+MGD it is possible to get more massive and compact objects than in the GR domain. Notwithstanding, the extreme value $u_{0}=4/9$ corresponding to the isotropic case in the GR arena remains as an upper bound, despite the matter distribution in this case contains local anisotropies. Of course, the possibility of overcoming $4/9$ in principle depends on the model, since $\alpha$ and $\lambda$ are subject to the metric functions and thermodynamic variables. In the critical case $u_{0}=4/9 \Rightarrow u>4/9$, then the black hole value $u_{BH}=1/2$ becomes automatically in the maximum bound for $u$. In the present case, it is evident that when $\lambda$ decreases $M$ and $u$ increase. However, it is clear that $\lambda$ is not a variable parameter, but it is necessary to take at least different orders of magnitude to  
establish what are the maximum values of $M$ and $u$ allowed by the theory.}
{On the other hand, all this analysis allows to elucidate how the Buchdahl limit is affected when one moves from GR framework to Rastall scenario (with and without MGD). As the expressions (\ref{news3}) and (\ref{news5}) shows the new contributions to this important quantity are non--trivial. Besides, these contributions strongly depend on the selected model, since it also determines the signature of the parameters $\alpha$ and $\lambda$ which are too much involved in the mass--radius ratio.}
It is worth mentioning that any order of magnitude outside the range considered for $\lambda$ and $\alpha$ in this opportunity, is ruled out. This is because the space parameter $\{A, B, C\}$ becomes imaginary. 

%%%%%%%%%%%%%%%%%%%%%%%%%
%%%%%%%%%%%%%%%%%%%%%%%%%%

\section{Concluding remarks}\label{sec9}
We extended gravitational decoupling via minimal geometric deformation approach into the Rastall gravity scenario. To illustrate how this methodology works in the background of Rastall gravity, the well known Tolman IV space--time describing a spherically symmetric and static perfect fluid sphere was analyzed. This model was already studied in the light of general relativity + minimal geometric deformation scheme \cite{r57} and in the arena of pure Rastall theory \cite{r45}. In both cases, the resulting model respects the general requirements in order to describe a well--behaved solution. 

Since Rastall theory of gravity contains an extra term which deviates the attention from general relativity behavior, in this work we have investigated the effects of this extra term and the possibility to obtaining compact structures which could serve to describe neutron or quark stars. Due to the presence of this additional term, the minimal coupling matter breaks down and in consequence, Bianchi's identities are violated (the conservation law of the energy--momentum tensor). This issue could in principle modified the junction condition mechanism as happened in f(R) gravity, for example. In this respect, we have discussed extensively how Rastall contribution remains inside the compact configuration, allowing the implementation of the most general matching conditions \i.e, the Israel-Darmois junction conditions \cite{r95,r96}. Moreover, as was pointed out by Rastall \cite{r1}, his proposal and Einstein theory share the same vacuum solution, the outer Schwarzschild space--time.

To translate the Tolman IV solution to an anisotropic domain in the Rastall framework, we have followed the same approach given in \cite{r57}. This approach consists in imposing some suitable conditions relating the thermodynamic seed observables with the corresponding components of the new sector \i.e, the $\theta$--sector. With this extra information at hand, the problem is closed because the decoupler function $f(r)$ and the full $\theta$--sector is determined. The methodology followed in this work in order to tackle the system of equations (\ref{eq34})--(\ref{eq35}) is known as the mimic constraints approach. Among all the possibilities the most common ones worked in the literature are: i) $p(r)=\theta^{r}_{r}$, ii) $\rho=\theta^{t}_{t}$, that is the $r-r$ component of the $\theta$--sector mimics the seed pressure $p(r)$ and the $t-t$ one mimics the seed energy--density $\rho(r)$. However, it should be noted that an adequate decoupler function $f(r)$ can be imposed in order to close the problem (for more details see \cite{r69,r70,r91}). The advantage of both proposals are evident. Regarding the first one, it allows to obtain the decoupler function $f(r)$ in an easy way. This is so because, one obtains after equating the corresponding field equation for $p(r)$ and $\theta^{r}_{r}$ is an algebraic equation (see Eq. (\ref{eq61})). The second choice does not lead to an algebraic equation, but to a first order differential equation (Eq. (\ref{eq75})). At this point, it is worth mentioning that in the case of general relativity + minimal geometric deformation, the Rastall contribution $F_{\lambda}$ is not there. So, obtain the decoupler function $f(r)$ is easier than our case, due to the Rastall piece $F_{\lambda}$ strongly depends on the metric potentials $\mu(r)$ and $\nu(r)$ (\ref{eq47}). So, when the $t-t$ component of the $\theta$--sector mimics the seed energy--density $\rho$, this additional term could introduce some mathematical complications. 

The emergence of Rastall term after impose the mimic constraints, is due to after splitting the system of equations (\ref{eq13})--(\ref{eq15}) by introducing the minimal geometric deformation (\ref{eq26}), the resulting seed sector (\ref{eq27})--(\ref{eq29}) was solved in order to express $p(r)$ and $\rho(r)$ in a separate way. The resulting expressions for $\rho(r)$ and $p(r)$ are (\ref{eq30}) and (\ref{eq31}) respectively that contain the usual Einstein terms and the Rastall contribution. This additional term is coupled to the field equations via a dimensionless constant $\lambda$, the so--called Rastall's parameter \cite{r1}.  Clearly, in the limit, $\lambda\rightarrow 0$ Einstein's gravity theory is recovered.  Thus, the fact to separate $p(r)$ and $\rho(r)$ introduces the Rastall contribution into the $\theta$--sector through the deformation function $f(r)$. Therefore,  the incidence of Rastall contributions are evident.  Since, not only Rastall's parameter $\lambda$ is affecting the dynamic of the solution, but also the extra geometrical terms. 

Mimic constraint methodology does not introduce new information into the problem. Because, these constraints are imposed at the level of the field equations, relating them after separate the system of equations (\ref{eq27})--(\ref{eq29}) by means of minimal geometric deformation approach. The consistency of these choices is reflected in the obtained solutions. Where in both cases, the evolution of thermodynamic parameters {reveals} an appropriate behavior as dictated by the basic requirements associated with the study of compact structures. Furthermore, the mimic constraint grasp plays an important role in some observational parameters such as the surface gravitational red-shift $z_{s}$. As it is well--known the surface gravitational red--shift relates the macro observables features of any compact configuration  \i.e, the mass and radius. In this respect, Ivanov studies \cite{r106} suggest that $z_{s}$ changes in magnitude when anisotropies are present in the material content. Moreover, B\"{o}hmer and Harko \cite{r107} discussed the effects on the compactness factor in the anisotropic matter distribution case. Notwithstanding, in the present study, the mimic constraint $p(r)=\theta^{r}_{r}$ does not modify the total mass of the compact object, it only redistributes the mass inside the stellar interior. Consequently, the compactness factor $u$ and surface gravitational red-shift $z_{s}$ remain unchanged, which makes it difficult to distinguish between an object whose material content is isotropic from an anisotropic one. In distinction with the case $\rho(r)=\theta^{t}_{t}$ where the total mass of the object is modified, therefore the observational implications are different.

To show how the anisotropic effects introduced by the $\theta$--sector work in the Rastall framework, we have revisited the behavior of the main salient features in the arena of general relativity, general relativity + gravitational decoupling minimally deformed and pure Rastall theory. In this concern we have fixed the space parameter to be $\{u,\alpha,\lambda,\}=\{{\mathrm{GR}}~\{0.2,0,0\};\mathrm{{GR}+MGD}~\{0.2,0.2,0\}$;
$\mathrm{RT}~\{0.2,0,-0.09\}$;  $\mathrm{RT+MGD} ~\{0.2,0.2,-0.09\}\}$ (RT means Rastall theory). From fig. \ref{fig1} (these plots correspond to $p(r)=\theta^{r}_{r}$ solution) it is clear that the RT+MGD radial and tangential pressures dominate the corresponding ones in the picture of GR and GR+MGD. Nonetheless, RT dominates all frames. Particularly, in comparing RT with RT+MGD, the final radial pressure in RT+MGD represents only a portion of the  pressure of the RT, indeed $p(r)_{(\mathrm{RT+MGD})}=\left(1-\alpha\right)p(r)_{(\mathrm{RT})}$. On the other hand, the final energy--density in the RT+MGD dominates all scenarios. So, by using $p(r)=\theta^{r}_{r}$ the final configuration is denser than {GR}, {GR}+MGD and RT. However, the increase in energy--density does not reflect a change in the total mass of the object (as discussed earlier). In Fig. \ref{fig2} (these plots correspond to $\rho(r)=\theta^{t}_{t}$ solution), the salient radial and tangential pressure in the RT+MGD picture dominate GR, {GR}+MGD and RT, what is more the salient energy--density also dominates over GR and {GR}+MGD and RT. Finally, both solutions present a positive anisotropy factor $\Delta$. In fact, this characteristic avoids the system to undergo unstable behavior. To back up this analysis and in order to provide some physical meaning to the mimic constraint approach, we have done a detailed study of the impact on the total mass contained by the fluid sphere and its associated mass--radius ratio. This study is displayed in Sec. \ref{sec7} and supported by Figs. \ref{fig3} and \ref{fig4}.\\
Moreover, by fixing the real observational data $M_{0}=1.97M_{\odot}$ and $R=11.7\ [\text{km}]$, corresponding to the millisecond pulsar PRS J1614--2230 \cite{demorest}, we have investigated the maximum total mass $M$ and compactness factor allowable for the theory under this particular model. To do so, we have bounded $\alpha$ and $\lambda$ by using (\ref{eq73}) and the salient physical variables associated to this constraint (we have not analyzed the situation when the restriction (\ref{eq60}) is imposed, because as was discussed this constraint does not modify the mass), determining that the maximum values for $M$ and $u$ are 2.919 and 0.367612, respectively. In table \ref{table1} are displayed different values for $M$ and $u$ corresponding to different orders of magnitude for $\lambda$ and $\alpha=10^{-1}$. Any other order of magnitude for $\lambda$ and $\alpha$ is discarded, since the parameter space that describes the geometry of the considered model \i.e, $\{A, B, C\}$ becomes imaginary. {In this concern, it should be highlighted that this is the first time that Buchdahl limit is explored within the framework of Rastall gravity theory (with and without MGD). As can be seen the non--trivial contributions coming from Rastall side (and also from MGD sector), provides a numerical data which is outside the scope of what is usually reported in the study of compact structures. However, the value $4/9$ still remains upper bound limit of the compactness. To overcome the value $4/9$ one needs to consider the GR compactness factor $u_{0}$ equal to this quantity, then the resulting numerical data for $u$ will be greater than $4/9$. Nevertheless, in that case the upper bound becomes the black hole one \i.e, $u_{BH}=1/2$.     }
From the above facts, {the} physical meaning for mimic constraints: $p(r)=\theta^r_r$ (\ref{eq60}) and $\rho(r)=\theta^t_t$ (\ref{eq73}) can be elucidated. In considering the first mimic constraint, it is clear that anisotropy enters the system by perturbing the seed pressure. This in principle suggests, that some of the macro physical observables of the system such as the total mass $M$ and related quantities such as the compactness factor $u$ and the surface gravitational red--shift $z_{s}$ are not altered. On the other hand, the second mimic constraint introduces anisotropy by disturbing the density of the seed solution. Clearly, the aforementioned observables and their related quantities are directly affected. In conclusion, if the anisotropy enters the system through a change in the original pressure of the system, from the physical point of view, certain quantities of the original system are preserved, while if it enters due to density disturbances, these  quantities are modified.

As a final remark, we
want to highlight two things. First, it is possible to obtain well behaved stellar interiors in the framework of Rastall gravity by using gravitational decoupling via minimal geometric deformation approach. 
The two families of solutions found in this work satisfy and share all the physical and mathematical properties required in the study of compact configurations, which serve to understand the behavior of real astrophysical objects such as neutron stars, for example. Second, it was found that Rastall theory is a promising scenario to study the existence of compact structures described by an anisotropic matter distribution, which results can be contrasted with the well--posed general relativity theory.
Moreover, as was discussed in Sec. \ref{sec2} among all the features that Rastall gravity theory shares with other non--conservative modified gravity theories \cite{bertolami,bertolami1,bertolami2,bertolami3}, it should be noted that in the cosmological scenario \cite{fabris,dutta} stands as a viable and promising proposal which, together with the study carried out in this work on stellar interiors, can potentially answer some of the unknowns that are open today.

\section{Acknowledgements}
S. K. Maurya and F. Tello-Ortiz acknowledge that this work is carried out under TRC project, grant No.-BFP/RGP /CBS/19/099, of the Sultanate of Oman. S. K. Maurya is thankful for continuous support and encouragement from the administration of University of Nizwa. F. Tello-Ortiz is partially supported by grant Fondecyt No. $1161192$, Chile. F. Tello-Ortiz thanks the financial support by the CONICYT PFCHA/DOCTORADO-NACIONAL/2019-21190856 and   
projects ANT-1856 and SEM 18-02 at the Universidad de Antofagasta, Chile. F. Tello-Ortiz thanks to Luciano Gabbanelli for fruitful discussions.


\begin{thebibliography}{9}

\bibitem{r1} P. Rastall, Phys. Rev. D, 6, 3357 (1972).

\bibitem{r2}
H. A Buchdhal, Mon. Not. Roy. Astron. Soc. 150, 1 (1970).

\bibitem{r3}
A.	Starobinsky, Phys. Lett. B 91, 99 (1980).

\bibitem{r4} S. D. Odintsov and V. K. Oikonomou
Phys. Rev. D 99, 064049 (2019).

\bibitem{r5} S. D. Odintsov and V. K. Oikonomou  Class. Quantum Grav. 36, 065008 (2019).

\bibitem{r6}
S. Capozziello, R. D’Agostino and O. Luongo, Gen. Rel. Grav. 51, 2 (2019).

\bibitem{r7} S. V. Chervon, A. V. Nikolaev, T. I. Mayorova, S. D. Odintsov and V. K. Oikonomou, Nuc. Phys. B 936, 597 (2018).

\bibitem{r8} S. Capozziello, S. Nojiria and S. D. Odintsov Phys. Lett. B 781, 99 (2018).

\bibitem{r9} S. Capozziello, C. A. Mantica and L. G. Molinari, arXiv:1810.03204 (2018).

\bibitem{r10} C. S. Santos, J. Santos, S. Capozziello and J. S. Alcaniz, Gen. Rel. Grav. 49, 50 (2017).

\bibitem{r11} A. V Astashenok et al, Class. Quantum Grav. 34, 205008 (2017).

\bibitem{r12} V. B. Jovanović, S. Capozziello, P. Jovanović and D. Borka, Phys. D. Univ. 14, 73 (2016).

\bibitem{r13} S. Capozziello, M. De Laurentis, R. Farinelli and S. D. Odintsov, Phys. Rev. D 93, 023501 (2016).

\bibitem{r14} A. V. Astashenok, S. Capozziello and S. D. Odintsov, Phys. Lett. B 742, 160 (2015).

\bibitem{r15} A.V. Astashenok, S. Capozziello and S.D. Odintsov,  Astrophys. Space. Sci. 355, 333 (2015). 

\bibitem{r16} A.~V.~Astashenok, S.~Capozziello and S.~D.~Odintsov, J. Cosmol. Astropart. Phys. 1312, 040 (2013).

\bibitem{r17} S. Capozziello, N. Frusciante and D. Vernieri, Gen. Rel. Grav. 44, 1881 (2012).

\bibitem{r18} S. Capozziello, M. De Laurentis, S. D. Odintsov and A. Stabile, Phys. Rev. D 83, 064004 (2011).

\bibitem{r19} S. Capozziello, M. De Laurentis and A. Stabile
Class. Quant. Grav. 27, 165008 (2010).

\bibitem{r20} S. Nojiri, S. D. Odintsov, D. Sáez-Gómez
Phys. Lett. B 681, 74 (2009).

\bibitem{r21} S. Capozziello, E. De Filippis and V. Salzano, Mont. Not. R. Astron. Soc. 394, 947 (2009).

\bibitem{r22} S. Capozziello, A. Stabile and A. Troisi,
Class. Quant. Grav. 25, 085004 (2008). 

\bibitem{r23} T. Harko, F. S. N. Lobo, S. Nojri and S. D. Odintsov, Phys. Rev. D 84, 024020 (2011).

\bibitem{r24}J. Wu , G. Li, T. Harko, and S. D. Liang,  Eur. Phys. J. C 78, 430 (2018).

\bibitem{r25}E. Barrientos , F. S. N. Lobo, S. Mendoza, G. J. Olmo, and D. Rubiera-Garcia,  Phys. Rev. D 97, 104041 (2018).

\bibitem{r26}D. Deb , B. K. Guha, F. Rahaman, and S. Ray, Phys. Rev. D 97, 084026 (2018).

\bibitem{r27} P. H. R. S. Moraes  and P. K. Sahoo,  Eur. Phys. J. C 77, 480 (2017).
 
\bibitem{r28}A. Das, S.  Ghosh, B. K. Guha, S. Das, F. Rahaman and S. Ray, Phys. Rev. D 95, 124011 (2017).

\bibitem{r29} P. H. R. S. Moraes, J. D. V.  Arba{\~n}il and M. Malheiro, J. Cosmol. Astropart. Phys., 06, 005 (2016).

\bibitem{r30}K. Koyama, Rep. Prog. Phys. 79, 046902 (2016).

\bibitem{r31}Z. Yousaf, K. Bamba and M. Z. Bhatti,  Phys. Rev. D 93, 064059 (2016).

\bibitem{r32}Z. Yousaf, K. Bamba and M. Z. Bhatti,  Phys. Rev. D 93, 124048 (2016).

\bibitem{r33}R. Zaregonbadi and M. Farhoudi, Gen. Relativ. Grav. 48, 142 (2016).

\bibitem{r34} E. H.Baffou, M. J. S. Houndjo, M. E. Rodrigues, A. V. Kpadonou and J. Tossa,  Phys. Rev. D 92, 084043  (2015).

\bibitem{r35}A. Alhamzawi  and  R. Alhamzawi,  Int. J. Mod. Phys. D 25, 1650020 (2015).

\bibitem{r36}H. Shabani and M. Farhoudi, Phys. Rev. D 90, 044031 (2014).

\bibitem{r37} T. Harko, Phys. Rev. D 90, 044067 (2014).

\bibitem{r38}S. Chakraborty, Gen. Relativ. Gravit. 45, 2039 (2013).

\bibitem{r39} K. Bamba, S. Capozziello, S. Nojiri and S. D. Odintsov,  Astrophys. Space Sci. 342, 155 (2012).

\bibitem{r40}S. Capozziello, M. De Laurentis, Phys. Rept. 509, 167 (2011).

\bibitem{r41} S. Capozziello and V. Faraoni, Beyond Einstein Gravity (Springer, New York) (2010).

\bibitem{r42} T. P. Sotiriou and  V. Faraoni,  Rev. Mod. Phys. 82, 451 (2010).

\bibitem{r43} E. H. Baffou, M. J. S. Houndjo, M. E. Rodrigues, A. V.  Kpadonou and J. Tossa,  Chin. J. Phys. 55, 467 (2007).

\bibitem{r44} S. Nojiri and S. D. Odintsov, Int. J. Geom. Meth. Mod. Phys. 4, 115 (2007).

\bibitem{m1}   S. K. Maurya, Abdelghani Errehymy, Debabrata Deb, Francisco Tello-Ortiz, Mohammed Daoud, Phys. Rev. D 100, 044014 (2019); arXiv:1907.10149.

\bibitem{m2}   S. K. Maurya, Ayan Banerjee, Francisco Tello-Ortiz, Physics of the Dark Universe  27, 100438 (2020);
 arXiv:1907.05209.

\bibitem{m3}   S. K. Maurya, Francisco Tello-Ortiz, Annals of Physics 414, 168070 (2020); arXiv:1906.11756.

\bibitem{m4}  Debabrata Deb, Sergei V. Ketov, S. K. Maurya, Maxim Khlopov, P. H. R. S. Moraes, Saibal Ray, Mon. Not.  R. Astr. Soc. 485, 5652 (2019).

\bibitem{r45} S. Hansraj, A. Banerjee and P. Channuie, Ann. Phys. 400, 320 (2019).

\bibitem{r46} C. Abbas and M. R. Shahzad, Eur. Phys. J. A 54, 211 (2018).

\bibitem{r47} C. Abbas and M. R. Shahzad, Astrophys. Space Sci. 364, 50 (2019).

\bibitem{r48} G. Abbas and M. R. Shahzad, Astrophys. Space Sci. 363, 251 (2018).

\bibitem{r49} Y. Heydarzade and F. Darabi, Phys. Lett. B 771, 365 (2017).

\bibitem{r50} Y. Heydarzade, H. Moradpour and F. Darabi, Can. J. Phys. 95, 1253
(2017).

\bibitem{r51} K. Bamba, A. Jawad, S. Rafique and H. Moradpour,Eur. Phys. J. C 78,
986 (2018).

\bibitem{r52}I. Lobo, P. Moradpour, J. P. M. Graca and I. G. Salako, Int. J. Mod. Phys.
D 27, 1850069 (2018).

\bibitem{r53} M. S. Ma and R. Zhao, Eur. Phys. J. C 77, 629 (2017).

\bibitem{r54} R. Kumar and S. G. Ghosh, Eur. Phys. J. C 78, 750 (2018).

\bibitem{r55} Z. Xu, X. Hou, X. Gong and J. Wangr, Eur. Phys. J. C 78, 513 (2018).

\bibitem{r56} J. Ovalle, Phys. Rev. D 95, 104019 (2017).

\bibitem{r57} J. Ovalle, R. Casadio, R. da Rocha and A. Sotomayor, Eur. Phys. J. C 78, 122 (2018).

\bibitem{r58} J. Ovalle, Laszl\'o A. Gergely and R. Casadio,  Class. Quantum Grav. 32, 045015 (2015).

\bibitem{r59} J. Ovalle,  Int. J. Mod. Phys. Conf. Ser. 41, 1660132 (2016).

\bibitem{r60} L. Randall and R. Sundrum,  Phys. Rev. Lett. 83, 3370 (1999).

\bibitem{r61} L. Randall and R. Sundrum,  Phys. Rev. Lett. 83, 4690 (1999). 

\bibitem{r62} J. Ovalle, Mod.Phys.Lett. A 23, 3247 (2008).

\bibitem{r63} J. Ovalle and F. Linares,   Phys.Rev. D 88, 104026 (2013). 

\bibitem{r64} J. Ovalle, F. Linares, A. Pasqua, A. Sotomayor,Class. Quantum Grav., 30, 175019 (2013).

\bibitem{r65}  R. Casadio, J. Ovalle, R. da Rocha,  Class. Quantum Grav. 30, 175019 (2014).

\bibitem{r66} R. Casadio, J. Ovalle and  R. da Rocha, Europhys. Lett. 110, 40003 (2015). 

\bibitem{r67} R. Casadio, J. Ovalle and R. da Rocha,  Class. Quantum Grav. 32, 215020 (2015). 

\bibitem{r68} J. Ovalle, R. ~Casadio and A. Sotomayor, Adv. High Energy Phys. 2017, 9 (2017).

\bibitem{r69}S. K. Maurya and F. Tello-Ortiz, Eur. Phys. J. C 79, 85 (2019). 

\bibitem{r70} E. Morales and F. Tello-Ortiz, Eur. Phys. J. C 78, 841 (2018).

\bibitem{r71} M. Estrada and F. Tello-Ortiz, Eur. Phys. J. Plus 133, 453 (2018).

\bibitem{r72} E. Morales and F. Tello-Ortiz, Eur. Phys. J. C 78, 618 (2018).

 
\bibitem{r73} L. Gabbanelli, A. Rinc\'on and C. Rubio, Eur. Phys. J. C 78, 370 (2018). 

\bibitem{r74} C. Las Heras and P. Le\'on, Fortsch. Phys. 66, 1800036 (2018).

\bibitem{r75} A. R. Graterol, Eur. Phys. J. Plus 133, 244 (2018).

\bibitem{r76} J. Ovalle and A. Sotomayor, Eur. Phys. J. Plus 133, 428 (2018).

\bibitem{r77} L. Gabbanelli, J. Ovalle, A. Sotomayor, Z. Stuchlik and R. Casadio,  Eur.Phys.J. C 79, 486 (2019).

\bibitem{r78} {S. Hensh, Z. Stuchl\'{i}k Eur.Phys.J. C 79, 834 (2019).}

\bibitem{r79}E. Contreras, A. Rinc\'on and P. Bargue\~{n}o, Eur. Phys. J. C, 79, 216 (2019).

\bibitem{r791} K. N. Singh, S. K. Maurya, M. K. Jasim, F. Rahaman, Eur. Phys. J. C 79, 851 (2019).

\bibitem{r792}F. Tello-Ortiz, S. K. Maurya, Y. Gomez-Leyton, Eur. Phys. J. C 80, 324 (2020).

\bibitem{r80}  J. Ovalle, R. Casadio, R. da Rocha
, A. Sotomayor and Z. Stuchlik, Eur. Phys. J. C 78, 960 (2018). 

\bibitem{r81} E. Contreras and P. Bargue\~{n}o, Eur. Phys. J. C 78, 558 (2018).

\bibitem{r82} E. Contreras and P. Bargue\~no, Eur. Phys. J. C, 78, 985 (2018).

\bibitem{r83} E. Contreras,  Eur. Phys. J. C, 78, 678 (2018).

\bibitem{r84} E. Contreras, Class. Quant. Grav. 36, 095004 (2019).

\bibitem{r85} G. Panotopoulos and A. Rinc\'on, 
Eur. Phys. J. C 78, 851 (2018).

\bibitem{r86} J. Ovalle, R. Casadio, R. Da Rocha, A. Sotomayor and Z. Stuchlik, EPL 124, 20004 (2018).

\bibitem{r87} {C. Las Heras and P. Le\'on  Eur. Phys. J. C 79, 990 (2019).}

\bibitem{r88} {J. Ovalle, C. Posada and Z. Stuchl\'ik, Class. Quantum Grav. 36, 205010 (2019).}

\bibitem{r89} J. Ovalle, Phys.Lett. B, 788, 213 (2019). 

\bibitem{r90} {E. Contreras and P. Bargue\~no, Class. Quantum Grav. 36, 215009 (2019).}

\bibitem{Maurya19} S. K. Maurya, Eur. Phys. J. C 79, 958 (2019)

\bibitem{r91}  M. Estrada, R. Prado, Eur.Phys.J.Plus 134, 168 (2019).

\bibitem{r92} M. Estrada, Eur. Phys. J. C 79, 918 (2019). 

\bibitem{r93} S. K. Maurya and F. Tello-Ortiz, Phys. Dark Univ. 27, 100442 (2020).

\bibitem{r94} F. X. L. Cede\~{n}o and E. Contreras, Phys. Dark Univ. 28, 100543 (2020).

\bibitem{visser} M. Visser, Phys. Lett. B, 782 83 (2018).

\bibitem{felice} A. De Felice and S. Tsujikawa, Living Rev. Rel. 13, 3 (2010).

\bibitem{epjc} F. Darabi, H. Moradpour, I. Licata, Y. Heydarzade and C. Corda, Eur. Phys. J. C 78, 25 (2018). 

\bibitem{moraes} W. A. G. De Moraes and A. F. Santos, \emph{ Gen. Rel. Grav.} \textbf{51}, 167 (2019).

\bibitem{moraes1} A. F. Santos and S. C. Ulhoa, \emph{Mod. Phys. Lett. A} \textbf{30}, 1550039 (2015).

\bibitem{bertolami} O. Bertolami, C. Bohmer, T. Harko and F. Lobo, Phys. Rev. D  75, 104016 (2007).

\bibitem{bertolami1} O. Bertolami and J. p\'{a}ramos, Phys. Rev. D 77, 084018 (2008).

\bibitem{bertolami2} O. Bertolami and A. Martins, Phy. Rev. D 85, 024012 (2012).

\bibitem{bertolami3} O. Bertolami and J. P\'{a}ramos, JCAP 1003, 009 (2010).

\bibitem{rawaf} A. S. Al-Rawaf and M. O. Taha, Phys. Lett. B366, 69 (1996).

\bibitem{hugo} H. Seeliger, Astro. Nach. 137, 129 (1895).

\bibitem{fabris} J. C. Fabris, O. F. Piattella, D. C. Rodrigues, C. E. M. Batista and M. H. Daouda,
 Int. J. Mod. Phys. Conf. Ser. 18, 67 (2012).
 
\bibitem{dutta} W. Khyllep and J. Dutta, Phys. Lett. B 797, 134796 (2019).

\bibitem{r105}
H. A.  Buchdahl,  Phys. Rev. D 116, 1027 (1959).

\bibitem{r107} C. G. B\"{o}hmer and T. Harko,  Class. Quantum Gravit. 23, 6479 (2006).

\bibitem{r95} W. Israel, Nuovo Cim. B 44, 1 (1966).

\bibitem{r96}  G. Darmois, M\'emorial des Sciences Mathematiques (Gauthier-Villars, Paris, 1927), Fasc. 25 (1927).

\bibitem{r99}
P. Musgrave and K. Lake, Class.Quant.Grav. 13, 1885 (1996).

\bibitem{r100} K. Lake, Phys. Rev. D 67, 104015 (2003) .

\bibitem{r101} K. Lake, Gen. Relativ. Gravit. 49, 134 (2017). 

\bibitem{r98}
J. M. M. Senovilla, Phys. Rev. D 88,  064015 (2013).

\bibitem{demorest} P. B. Demorest, T. Pennucci, S. M. Ransom, M. S. E. Roberts and J. W. T Hessels, \emph{Nat.} \textbf{467}, 1081 (2010).

\bibitem{linares} F. X. Linares, M. A. Garcia--Aspeitia and L. A. Ure\~{n}a--Lopez, \emph{Phys. Rev. D} \textbf{92}, 024037 (2015).

\bibitem{r102}
H. Abreu, H. Hernandez, L. A. N\'{u}\~{n}ez, Class. Quantum Gravit. 24, 4631 (2007).

\bibitem{r103}
L. Herrera, Phys. Lett. A 165, 206 (1992).

\bibitem{r104} M. K. Gokhroo and A. L. Mehra,  Gen. Rel. Grav. 26, 75 (1994).

\bibitem{r106} B.V. Ivanov, Phys. Rev. D 65, 104011 (2002).


\end{thebibliography}
\end{document}